\documentclass[twocolumn, twocolappendix]{aastex631}

\newcommand\kms{\mathrm{km}\,\mathrm{s}^{-1}}
\newcommand\lu{$10^3\,\mathrm{kpc\,km\,s}^{-1}$}
\newcommand\lms{$L_\mathrm{MS}$}
\newcommand\bms{$B_\mathrm{MS}$}
\newcommand\feh{\mathrm{[Fe/H]}}

\newcommand\nmss{13}
\newcommand\dgrad{$-0.5 \pm 0.2$}
\newcommand\gasmass{$(5.2 \pm 1.2) \times 10^8\,M_\odot$}

\usepackage{CJKutf8}

\shorttitle{The Magellanic Stellar Stream}
\shortauthors{Chandra et al.}

\graphicspath{{./}{fig/}}

\begin{document}

\defcitealias{Zaritsky2020}{Z20}

\title{Discovery of the Magellanic Stellar Stream Out to 100 Kiloparsecs}

\author[0000-0002-0572-8012]{Vedant~Chandra}
\affiliation{Center for Astrophysics $\mid$ Harvard \& Smithsonian, 60 Garden St, Cambridge, MA 02138, USA}

\author[0000-0003-3997-5705]{Rohan~P.~Naidu}
\altaffiliation{NASA Hubble Fellow}
\affiliation{MIT Kavli Institute for Astrophysics and Space Research, 77 Massachusetts Ave., Cambridge, MA 02139, USA}

\author[0000-0002-1590-8551]{Charlie~Conroy}
\affiliation{Center for Astrophysics $\mid$ Harvard \& Smithsonian, 60 Garden St, Cambridge, MA 02138, USA}

\author[0000-0002-7846-9787]{Ana~Bonaca}
\affiliation{The Observatories of the Carnegie Institution for Science, 813 Santa Barbara Street, Pasadena, CA 91101, USA}

\author[0000-0002-5177-727X]{Dennis~Zaritsky}
\affiliation{Steward Observatory and Department of Astronomy, University of Arizona, Tucson, AZ 85721, USA}

\author[0000-0002-1617-8917]{Phillip~A.~Cargile}
\affiliation{Center for Astrophysics $\mid$ Harvard \& Smithsonian, 60 Garden St, Cambridge, MA 02138, USA}

\author[0000-0003-2352-3202]{Nelson~Caldwell}
\affiliation{Center for Astrophysics $\mid$ Harvard \& Smithsonian, 60 Garden St, Cambridge, MA 02138, USA}

\author[0000-0002-9280-7594]{Benjamin~D.~Johnson}
\affiliation{Center for Astrophysics $\mid$ Harvard \& Smithsonian, 60 Garden St, Cambridge, MA 02138, USA}

\author[0000-0002-6800-5778]{Jiwon~Jesse~Han}
\affiliation{Center for Astrophysics $\mid$ Harvard \& Smithsonian, 60 Garden St, Cambridge, MA 02138, USA}

\author[0000-0001-5082-9536]{Yuan-Sen~Ting \begin{CJK*}{UTF8}{gbsn}(丁源森)\end{CJK*}}
\affiliation{Research School of Astronomy \& Astrophysics, Australian National University, Cotter Road, Weston, ACT 2611, Australia}
\affiliation{School of Computing, Australian National University, Acton ACT 2601, Australia}
\affiliation{Department of Astronomy, The Ohio State University, Columbus, OH 43210, USA}

\correspondingauthor{Vedant Chandra}
\email{vedant.chandra@cfa.harvard.edu}

\begin{abstract}
\noindent
The Magellanic Stream (MS) --- an enormous ribbon of gas spanning $140^\circ$ of the southern sky trailing the Magellanic Clouds --- has been exquisitely mapped in the five decades since its discovery. 
However, despite concerted efforts, no stellar counterpart to the MS has been conclusively identified. 
This stellar stream would reveal the distance and 6D kinematics of the MS, constraining its formation and the past orbital history of the Clouds. 
We have been conducting a spectroscopic survey of the most distant and luminous red giant stars in the Galactic outskirts. 
From this dataset, we have discovered a prominent population of \nmss{} stars matching the extreme angular momentum of the Clouds, spanning up to $100^\circ$ along the MS at distances of $60-120$~kpc. 
Furthermore, these kinemetically-selected stars lie along a [$\alpha$/Fe]-deficient track in chemical space from $-2.5 < \mathrm{[Fe/H]} < -0.5$, consistent with their formation in the Clouds themselves.
We identify these stars as high-confidence members of the Magellanic Stellar Stream. 
Half of these stars are metal-rich and closely follow the gaseous
MS, whereas the other half are more scattered and metal-poor. 
We argue that the metal-rich stream is the recently-formed tidal counterpart to the MS, and speculate that the metal-poor population was thrown out of the SMC outskirts during an earlier interaction between the Clouds.
The Magellanic Stellar Stream provides a strong set of constraints --- distances, 6D kinematics, and birth locations --- that will guide future simulations towards unveiling the detailed history of the Clouds. 

\end{abstract}


\section{Introduction} \label{sec:intro}

The Magellanic Stream (MS) is an immense gaseous structure encircling the Milky Way, spanning over $140^\circ$ of the southern sky \citep[see][for a comprehensive review]{D'Onghia2016}. 
Initially discovered via HI gas, all-sky radio surveys have revealed the full extent of the stream emerging in the wake of the Magellanic Clouds (hereafter `Clouds'; see \citealt{Dennefeld2020} for a discussion about this name), and also in a leading arm ahead of the Clouds \citep{Dieter1965, Wannier1972, Mathewson1974, Putman1998, Lu1998, Venzmer2012, For2013}. 
The MS is bifurcated into two interwoven filaments separable in velocity and chemical composition, with one component originating from each of the Clouds \citep{Cohen1982, Morras1983, Putman2000, Putman2003b, Nidever2008, Fox2013, Richter2013}. 

Despite extensive observations and simulations of the gaseous MS, its origin has remained enigmatic for decades. 
In particular, the relative contribution from two major formation processes --- tidal disruption and ram-pressure stripping --- remains uncertain. 
The tidal disruption model suggests that the MS primarily formed through past gravitational interactions between the Clouds, and predicts several observed features of the MS like its leading arm  \citep{Fujimoto1976, Lin1977, Murai1980, Yoshizawa2003, Connors2006, Besla2012, Besla2013, Guglielmo2014}. 
The ram-pressure model forms the MS predominantly via hydrodynamical drag on the Clouds' gas, with both the MW and LMC's coronal gas playing a key role \citep{Meurer1985, Moore1994, Murali2000, Diaz2011a, Wang2019, Lucchini2020, Krishnarao2022}. 
In practice, a combination of both models is likely required to explain the concert of observed stream properties.

A key source of uncertainty is the past orbit of the Clouds, which is significantly affected by hydrodynamical influences that are challenging to model \citep{Lucchini2021a}. 
Regardless, proper motion measurements of the Clouds have revealed that they are likely on their first pericentric passage around the Milky Way as a loosely bound pair \citep[][although see \citealt{Vasiliev2023b}]{Besla2007,D'Onghia2008,Nichols2011,Kallivayalil2013,Zivick2018}. 
This has broad implications for our own Milky Way, due to the LMC's outsize influence on the Milky Way's stellar and dark matter halo (e.g., \citealt{Erkal2020, Erkal2021, Vasiliev2021a, Conroy2021}; see \citealt{Vasiliev2023} for a recent review). 
The MS can likewise teach us about our own Galaxy, from measuring the MW's mass \citep[e.g.,][]{Craig2022} to revealing recent Seyfert activity in the MW's supermassive black hole \citep{Bland-Hawthorn2013}. 
Constraints from the MS are highly sensitive to its distance, which has historically been difficult to pin down in the absence of a stellar component. 

Models of the MS predict that there should be a tidally-stripped stellar counterpart that traces the gaseous stream \citep[e.g.,][]{Lin1995,Gardiner1996,Connors2006,Ruzicka2010,Diaz2011a, Diaz2012,Besla2012,Besla2013}. 
The existence and characteristics of this `Magellanic Stellar Stream' (MSS) would place strong constraints on the distance and kinematics of the gaseous MS, and consequently the past interaction history of the Clouds. 
Furthermore, in order to quantify the influence of the LMC on the MW's outer stellar halo, unmixed substructure like the MSS must be carefully modeled or removed \citep{Garavito-Camargo2019, Garavito-Camargo2021, Garavito-Camargo2021b}.
Finally, the MSS and MS together would sensitively trace the joint gravitational potential of the MW and Clouds, enabling measurements of their dark matter distribution \citep[e.g.,][]{Vasiliev2021a, Lilleengen2023}. 

However, despite concerted efforts over five decades, no unambiguous extended MSS has been identified to date \citep[e.g.,][]{Recillas-Cruz1982,Brueck1983,Guhathakurta1998}. 
Stripped stars have been identified in the Magellanic Bridge feature connecting the Clouds, which is thought to be the imprint of a past direct collision \citep{Irwin1990, Demers1991, Harris2007, Bagheri2013, Noel2013, Skowron2014}.
\cite{Belokurov2016} used BHB stars from the Dark Energy Survey to find stream-like protrusions extending away from the Clouds at $50-80$~kpc, including a $20^\circ$ long stream aligned with the gaseous MS. 
\cite{Petersen2022} present evidence of RR Lyrae (RRL) stars in the leading arm of the MS at $\approx 50$~kpc, but a definitive association is challenging without radial velocities and abundances. 

Recently, \citet[][hereafter \citetalias{Zaritsky2020}]{Zaritsky2020} presented 15 stars in the H3 Survey that are spatially coincident with the trailing tip of the MS, lie $\approx 50$~kpc from the Sun, and have a similar Galactocentric radial velocities to the MS HI gas in that region. 
They identified these stars as originating from the Clouds based on their highly negative Galactocentric radial velocities and cold velocity dispersion. 
However, a hallmark feature of the orbital trajectory of the Clouds is a large negative angular momentum around the Galactocentric X axis ($L_\mathrm{X}$, see Figure~\ref{fig:simL}), which the \citetalias{Zaritsky2020} stars puzzlingly lack.

As the above examples demonstrate, it is challenging to definitively associate stars to the MS without 6D kinematics (proper motions and radial velocities) and chemical abundances. 
In the inner halo ($\lesssim50$ kpc), it is only with exquisite \textit{Gaia} astrometry and ground-based spectroscopic surveys that various stellar streams, globular clusters, and field stars have been traced back to distinct dwarf galaxies \citep[e.g.,][]{Helmi2020, Kruijssen2020, Naidu2020, Bonaca2020a, Malhan2022a}. 
Here we seek to apply a similar approach to the outer halo, and to the quest for the MSS.

We have been executing a spectroscopic survey of the Milky Way's mostly uncharted outskirts beyond $\gtrsim 50$~kpc. 
This survey complements existing halo surveys like the H3 Survey \citep{Conroy2019b} with its focus on the most distant stars in the Galaxy. 
In this work, we present the first results from our survey: \nmss{} high-confidence members of the MSS beyond $60$~kpc, spanning over a hundred degrees along the gaseous MS. 
These stars are kinematic outliers in our spectroscopic sample, with past orbital trajectories leading back to the Clouds. 
Furthermore, they have chemical abundances consistent with their formation in the Clouds themselves. 

We describe our data collection and analysis in $\S$\ref{data}, and present the MSS in $\S$\ref{mss}. 
We discuss the implications of our results and conclude in $\S$\ref{discuss}. 
Throughout this work, we use `MS' to refer to the gaseous stream, and `MSS' to refer to the stellar counterpart we present here.

\section{Data \& Analysis}\label{data}

In this section we briefly describe our ongoing spectroscopic survey of luminous red giant branch (RGB) stars in the $\gtrsim 50$~kpc outer halo. This work presents the first scientific results from our survey, with other results currently in preparation. 

\subsection{Target Selection}\label{data:targets}

We first assemble a sample of plausible RGB stars using \textit{Gaia}~DR3 astrometry and unWISE infrared photometry \citep{Mainzer2014,Schlafly2019,GaiaCollaboration2021,Lindegren2021,GaiaCollaboration2022}. 
The selection procedure for our parent sample is described in \cite{Chandra2023}. 
Briefly, we initially query the \textit{Gaia} DR3 catalog for stars with $\varpi < 0.4$~mas, $\mu < 5$~mas/yr, and $\mid b \mid > 20^\circ$ to remove obvious nearby dwarfs and mask out the MW disk. 
We derive a significance statistic $\chi_{\mathrm{\varpi}}$ that encodes how many standard deviations lie between the observed \textit{Gaia} parallax and the predicted parallax for a dwarf with the same apparent magnitude and color. 
Using stars from the H3 Spectroscopic Survey \citep{Conroy2019b} as a guide, \cite{Chandra2023} find that removing stars with $\chi_{\mathrm{\varpi}} < 2$ results in a $\approx 90\%$ pure sample of stars with $\log{g} < 3.5$. 
We further purify this sample by applying a broad color cut in the $\left(\text{BP}-\text{RP},\text{RP}-\text{W1}\right)$ color space to remove dwarfs based on their \textit{WISE} infrared colors \citep{Conroy2018,Conroy2021}, which increases the sample purity to $\gtrsim 95\%$.

We derive approximate distances to these presumed giants using a 10~Gyr MIST isochrone \citep{Choi2016} with [Fe/H]\,$=-1.2$, matching the peak of the H3 Survey's metallicity distribution beyond $30$~kpc \citep{Conroy2019a}.
Although we only use these photometric distances for target selection, we note that their precision has been validated at the $20\%$ level compared to clusters, dwarf galaxies, and stars from the H3 Survey \citep{Conroy2021}.
We mask out stars that lie within $1^\circ$ of known dwarf galaxies and globular clusters from the catalogs of \cite{McConnachie2012} and \citep{Baumgardt2020} respectively. 
We also remove stars in the northern Galactic hemisphere that are beyond $d > 50$~kpc and lie within $20^\circ$ of the Sagittarius Stream. 
Our final sample of targets contains $\approx 400$ stars with implied distances $\gtrsim 100$~kpc. 
This is the primary target sample for our spectroscopic survey. 

For the present work, we additionally targeted stars with $50 \lesssim d \lesssim 100$~kpc that lie near the Pisces Plume overdensity (see Figure~\ref{fig:allsky}). 
The Pisces Plume is thought to mainly be composed of field halo stars that have been dragged by the dynamical friction wake of the LMC \citep{Belokurov2019a, Conroy2021}. 
However, it shares an on-sky location with the gaseous MS, and consequently the Pisces Plume could contain stars from the MSS, which would stand out due to their extreme angular momenta and chemical abundances. 
Since these stars lie closer than our main $d > 100$~kpc sample, all are brighter than $G < 17.65$ and consequently have low-resolution `XP' spectra available from \textit{Gaia} DR3 \citep{GaiaCollaboration2022, DeAngeli2022, Montegriffo2022a}. 
We utilized the \textit{Gaia} XP K giant catalog of \cite{Chandra2023} to select stars in the Plume, since the XP metallicities enable more precise distance estimates than the color--isochrone methodology outlined above. 
There were a total of $\approx 100$~stars in this Pisces Plume selection, of which we randomly observed 45~stars. 

\subsection{Observations and Analysis}

We have been executing a spectroscopic survey of the above selected giants with the Magellan Echellete Spectrograph (MagE; \citealt{Marshall2008}) on the 6.5m Magellan Baade Telescope at Las Campanas Observatory (PIs: Chandra \& Naidu). 
The stars are typically between $17 \lesssim G \lesssim 18.6$, and we utilize magnitude-dependent exposure times ranging from $10-40$~minutes per star to achieve signal-to-noise ratio (SNR)~$\approx 10~\mathrm{pixel}^{-1}$ at $5100$~\AA. 
We take a single ThAr arc exposure after each science exposure to ensure a reliable wavelength calibration.

We reduce our data with a fully automated pipeline\footnote{\url{https://github.com/vedantchandra/merlin}} built around the \texttt{PypeIt} utility \citep{pypeit:joss_pub,pypeit:zenodo}. 
Spectra are flat-fielded, wavelength-calibrated with the paired arc exposure, and then optimally extracted by \texttt{PypeIt}. 

We derive stellar parameters for our stars using the fully Bayesian \texttt{MINESweeper} fitting routine \citep{Cargile2020}. 
We fit the Mg triplet region of the MagE spectra (from 4800--5500~\AA), where our linelists are best-calibrated. 
We also include all archival broadband photometry in the likelihood, along with the \textit{Gaia} parallax \citep{Fukugita1996, Gunn1998, Skrutskie2006,  Mainzer2014, Chambers2016, GaiaCollaboration2021, GaiaCollaboration2022}. 
\texttt{MINESweeper} compares these observables to synthetic Kurucz model spectra, and additionally constrains the models to lie on MIST isochrones \citep{Kurucz1970, Kurucz1981, Choi2016}.

For our main survey catalogs we utilize a broad Gaussian prior on the stellar ages --- centred at 10~Gyr with a 4~Gyr dispersion, truncated between $4-14$~Gyr --- informed by the age distribution of more nearby halo stars \citep[e.g.,][]{Bonaca2020}. 
For the present search for MSS stars, we adopt a uniform prior between $4-14$~Gyr on the stellar ages, since the Clouds have an extended star formation history \citep[e.g.,][]{Nidever2020}. 
Furthermore, the age distribution of the outer halo is quite unconstrained. 
As long as very young (age $< 4$~Gyr) solutions are excluded, the choice of age prior affects isochrone distances at the $\lesssim 10\%$ level. 
Our main results are not significantly altered by the choice of prior, i.e., we can recover the MSS stars regardless of the assumed prior. 

The posterior distribution of stellar parameters is sampled with \texttt{dynesty} \citep{Speagle2020}, producing measurements of the radial velocity $v_\mathrm{r}$, effective temperature $T_\mathrm{eff}$, surface gravity $\log{g}$, metallicity [Fe/H], [$\alpha$/Fe] abundance, and heliocentric distance. 
Repeat observations of bright radial velocity standard stars --- HIP4148 and HIP22787 --- over multiple nights indicate that our systematic RV precision floor is $\approx 1\,\kms$, and the statistical radial velocity uncertainty reported by \texttt{MINESweeper} is typically $\approx 0.5\,\kms$ for our main survey sample. For more details on the implementation and validation of \texttt{MINESweeper}, we defer to \cite{Cargile2020}. Coordinates are transformed to a right-handed Galactocentric frame with a solar position $\mathbf{x}_\odot = (-8.12, 0.00, 0.02)$~kpc, and solar velocity $\mathbf{v}_\odot = (12.9, 245.6, 7.8)$~$\mathrm{km\,s^{-1}}$ \citep{Reid2004,Drimmel2018,GravityCollaboration2018}. 

\begin{figure}
    \centering
    \includegraphics[width=\columnwidth]{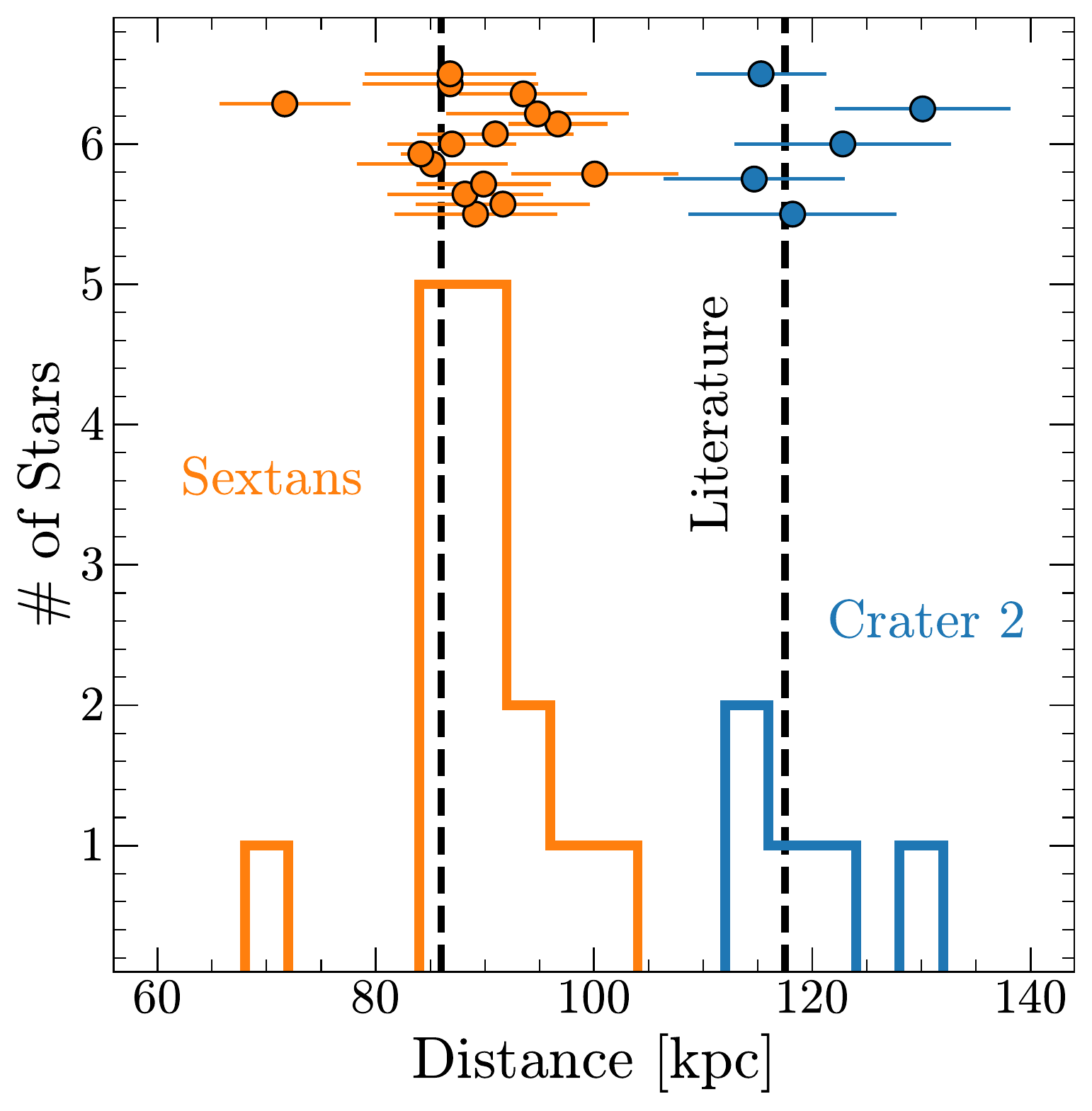}
    \caption{Validating the spectroscopic distance scale of our survey with stars from the Sextans and Crater~2 dwarf galaxies. We recover literature distances within $1\sigma$ in the mean, and the statistical distance uncertainties are well-calibrated.}
    \label{fig:dval}
\end{figure}
 
To test both the absolute and relative scale of our spectroscopic distances, we observed giants from our K giant catalog belonging to the Sextans ($86 \pm 1$~kpc, \citealt{Irwin1990, Munoz2018b}) and Crater~2 ($117 \pm 1$~kpc, \citealt{Torrealba2016b}) dwarf galaxies. 
We observed 15 K giants in Sextans, and 5 stars in Crater~2 that were shared between our K giant catalog and the spectroscopic sample of \cite{Ji2021}. 
Our resulting spectroscopic distances are shown in Figure~\ref{fig:dval}, with literature distances overlaid. 
Our inverse-variance-weighted mean spectroscopic distance to Sextans (Crater~2) is $86 \pm 1$~kpc ($119 \pm 3$~kpc), validating the absolute distance scale of the survey.
Furthermore, the distance uncertainties appear to be well-calibrated, with 60\% (100\%) of the stars lying within $1\sigma$ ($2.5\sigma$) of the literature distances. 

To date, we have observed $225$~stars in this survey, of which $191$ are spectroscopically confirmed to lie beyond 50~kpc, and $53$ are beyond 100~kpc. This is already the largest dataset of stars with precise abundances and 6D phase-space information beyond 100~kpc. Broader results from this dataset --- along with a public data release --- will be presented in forthcoming publications. For the present work, we select $191$ stars with spectroscopic $d > 50$~kpc, targeted from both selections described in $\S$\ref{data:targets}: photometry + parallax selected stars estimated to lie beyond $\approx 100$~kpc, and stars from \cite{Chandra2023} beyond $\approx 50$~kpc that lie within the Pisces Plume overdensity. 

\section{The Magellanic Stellar Stream}\label{mss}

\subsection{Selection with Angular Momenta}\label{sec:mss_sel}

\begin{figure*}
    \centering
    \includegraphics[width=0.45\textwidth]{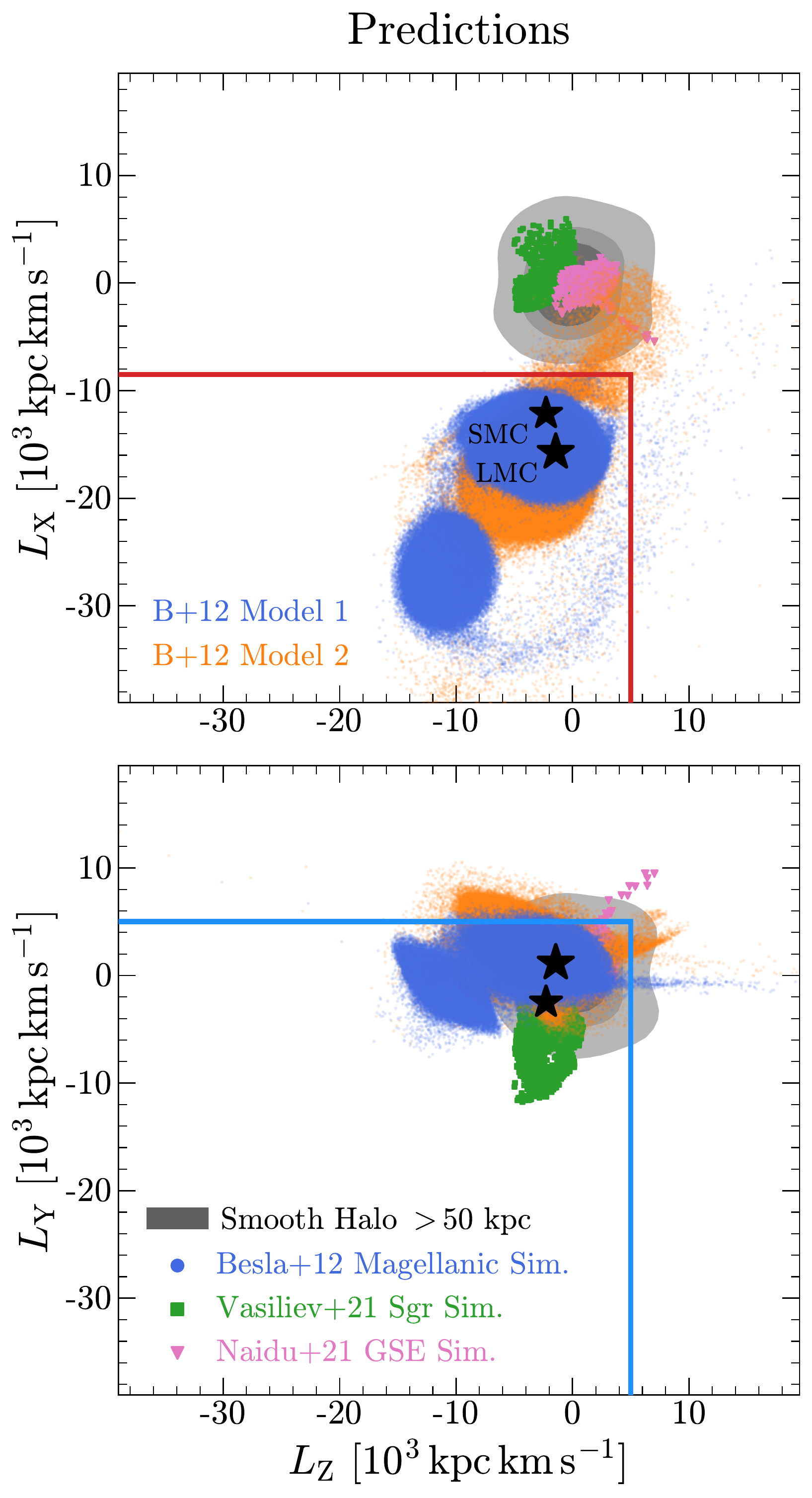}
    \includegraphics[width=0.45\textwidth]{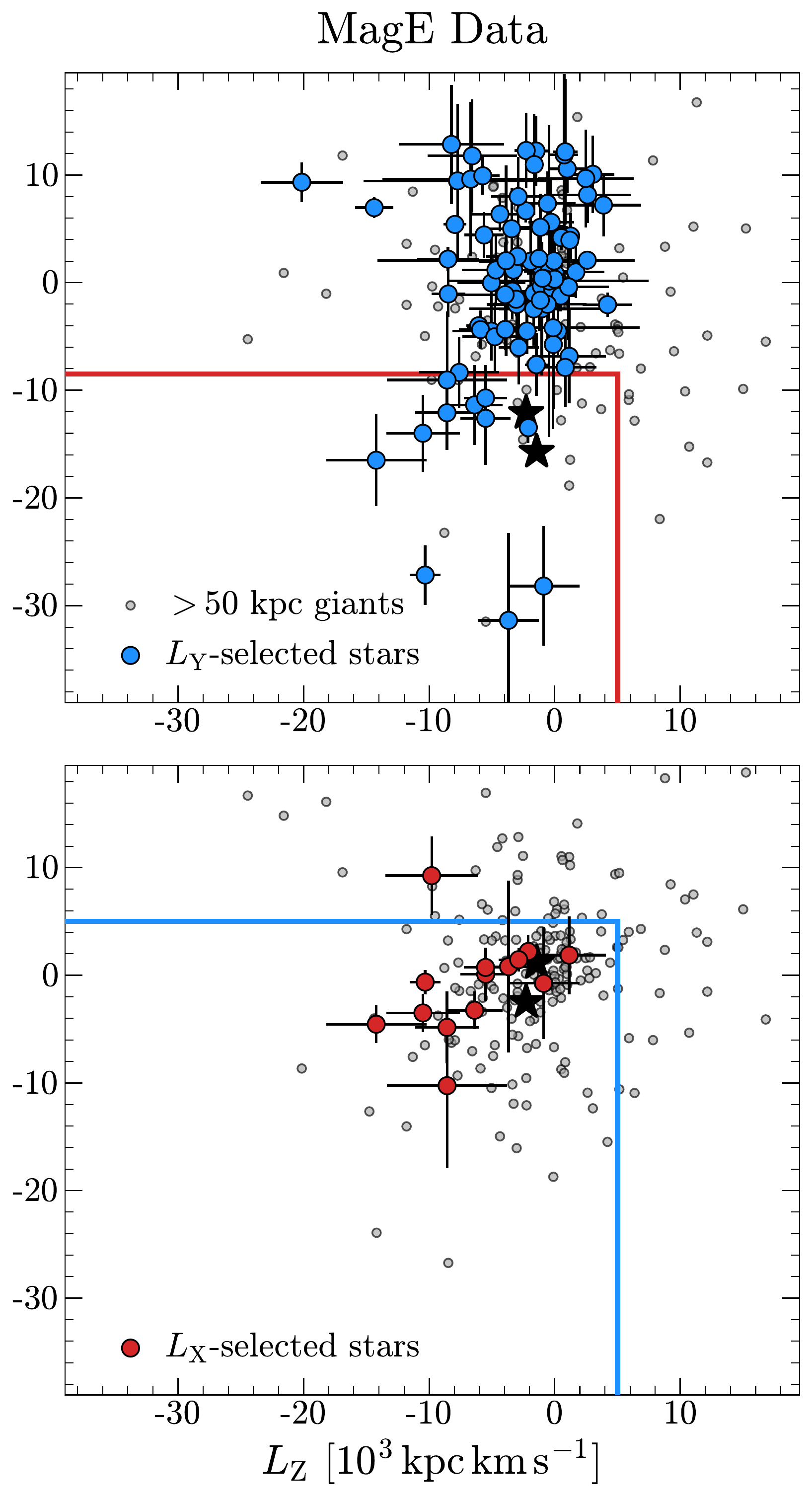}
     \caption{Selecting stars associated with the Clouds in angular momentum space. 
     The left panels show predictions from existing data and simulations. 
     Grey contours indicate the 3$\sigma$ distribution of stars beyond 50~kpc in the smooth mock halo of \cite{Robin2012}. 
     We show simulation particles beyond $30$~kpc the three largest contributors to the outer halo: the Clouds and MSS (\citealt{Besla2012,Besla2013}, blue and orange), the Sagittarius stream (\citealt{Vasiliev2021a}, green), and the Gaia Sausage Enceladus merger (\citealt{Naidu2021}, pink). 
     Black stars indicate the observed mean angular momenta of the Clouds. 
     The right panels show data from our MagE survey beyond 50~kpc, along with the corresponding red (blue) selection made in the $L_\mathrm{X}$ ($L_\mathrm{Y}$) space.
     The combination of $L_\mathrm{X}$-$L_\mathrm{Y}$-$L_\mathrm{Z}$ cuts isolates stars moving with the Clouds. 
     }
    \label{fig:simL}
\end{figure*}

\begin{figure*}
    \centering
    \includegraphics[width=\textwidth]{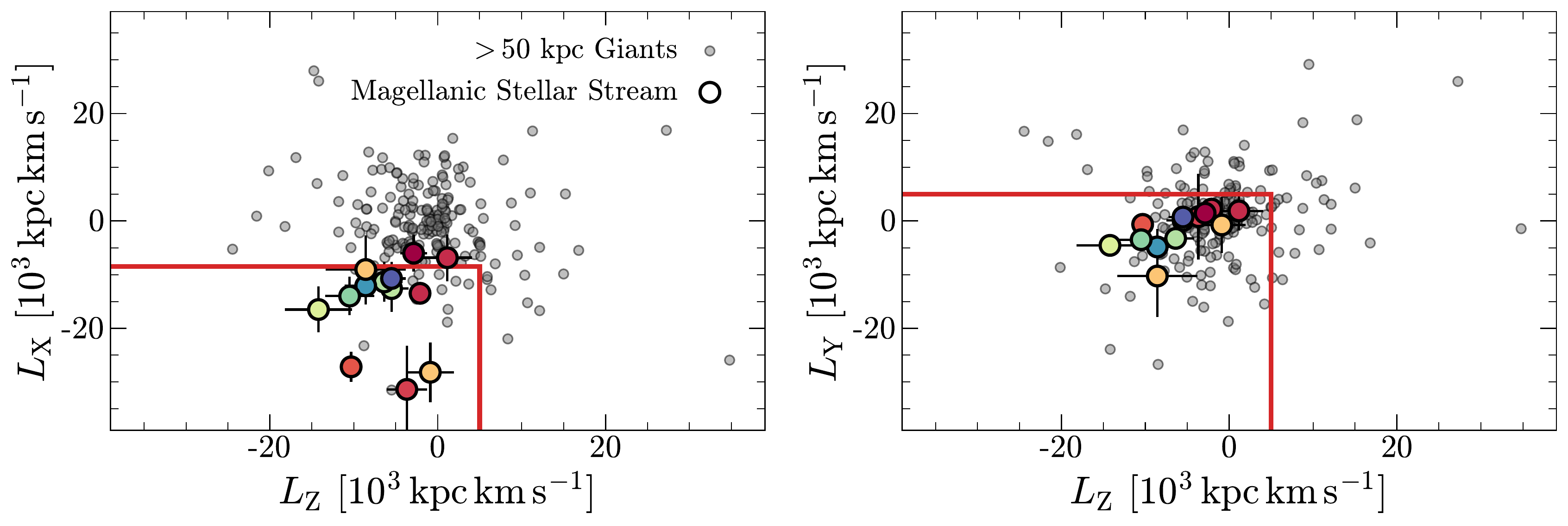}
    \includegraphics[height=6.5cm]{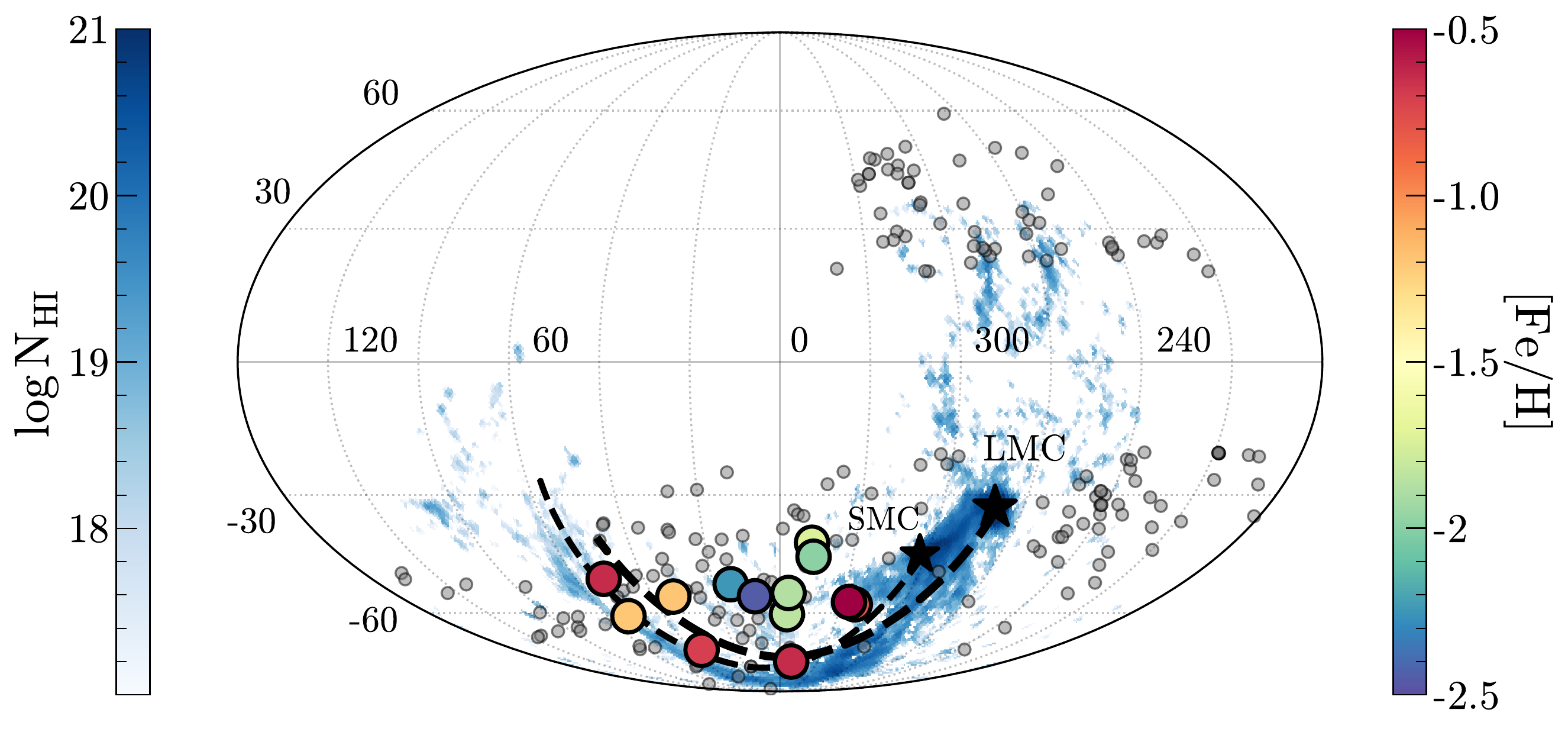}
    \caption{Selection of the Magellanic Stellar Stream from our survey data. Top: 191 stars beyond $50$~kpc in angular momentum space, with our kinematically-selected MSS stars colored by metallicity. The MSS stands out as a prominent tail of stars with highly negative $L_{\mathrm{X}}$, and we select stars within $1\sigma$ of the adopted $L_\mathrm{X}$ selection boundary. Bottom: \ion{H}{1} column density of the MS, with our identified MSS stars overlaid. The dashed lines trace the past $500$~Myr orbit of the Clouds \citep{Patel2020}.
    }
    \label{fig:mss}
\end{figure*}

The LMC and SMC are likely on first infall, and consequently possess high angular momentum that is yet to dissipate via dynamical friction \citep{Kallivayalil2006, Besla2007, Kallivayalil2013}. The angular momentum vector of the Clouds lies almost parallel to the Galactocentric X axis, causing the Clouds to have highly negative angular momenta in the $L_\mathrm{X}$ component: $\approx -15.7~(-12.1)\,\times$\,\lu{} for the LMC (SMC). 
One would expect that stellar debris stripped from the Clouds will have similar $L_\mathrm{X}$ --- or even more negative, since the angular momentum decays as the Clouds fall into the MW. 

The left panels of Figure~\ref{fig:simL} illustrate predicted angular momenta for various contributors to the outer halo beyond $30$~kpc, including a mock `smooth' halo, the Clouds and the MSS, the Sagittarius Stream, and the Gaia Sausage Enceladus (GSE) merger \citep{Robin2012, Besla2012,Vasiliev2021a,Naidu2021}. 
The MS Model 1 from \cite{Besla2012} is designed to best match the kinematics of the gaseous MS, whereas Model 2 is a better match to the present-day kinematics of the Clouds themselves. 
Figure~\ref{fig:simL} demonstrates the power of angular momenta --- chiefly $L_\mathrm{X}$ --- to isolate debris originating from the Clouds. 

In the right panels of Figure~\ref{fig:simL} we apply this angular momentum selection methodology to our survey sample of $> 50$~kpc giants. 
The top-right panel of Figure~\ref{fig:simL} shows our 191 MagE survey stars beyond $50$~kpc in $L_\mathrm{Z}-L_\mathrm{X}$ space. 
A prominent tail of stars with with modest $| L_\mathrm{Z} |$ but highly negative $L_\mathrm{X}$ is immediately apparent, with few corresponding analogs at positive $L_\mathrm{X}$. 
We perform a selection of $L_\mathrm{X}< 8.5~\times$~\lu{} and \lms{}\,$ < -15^\circ$ to isolate plausible MSS stars, also including stars with $L_\mathrm{X}$ within $1\sigma$ of the selection boundary.
The boundary is chosen to select stars with $L_\mathrm{X}$ plausibly as or more extreme than the Clouds.
We manually excise one star (\textit{Gaia} DR3 6538374163768839424) that lies very close to these selection boundaries but has very large angular momentum errors, as well as two stars with large proper motion errors and $V_\mathrm{GSR} \gtrsim 0~\kms{}$. 
Another cut is applied to remove one energetic field star with highly positive $L_\mathrm{Y}$ (bottom-right panel of Figure~\ref{fig:simL}), since the Clouds have modest $L_\mathrm{Y} \lesssim 1\,\times$~\lu{}. 
The selected region of negative $L_\mathrm{Y}$ also contains the Sagittarius stream, but we expect Sagittarius stars to be confidently removed by the $L_\mathrm{X}$ cut. 
The cuts presented in this section are admittedly fine-tuned to produce a pure sample of MSS candidates, and more detailed modeling may isolate further MSS candidates from our sample.

\subsection{The Stellar and Gaseous Magellanic Streams}

The combination of angular momentum cuts described in Figure~\ref{fig:simL} (and reproduced in the top panels of Figure~\ref{fig:mss}) efficiently selects stars moving in the same high-momentum orbital plane as the Clouds. 
This kinematic selection yields \nmss{} stars that trail the past orbit of the Clouds on-sky (Figure~\ref{fig:mss}, bottom). 
We show the on-sky distribution of our entire MagE spectroscopic sample in grey.
In the background we display the log-normalized \ion{H}{1} column density of the gaseous MS, with the MW's \ion{H}{1} contribution subtracted out \citep{Nidever2010}.

\begin{figure*}
    \centering
    \includegraphics[width=\textwidth]{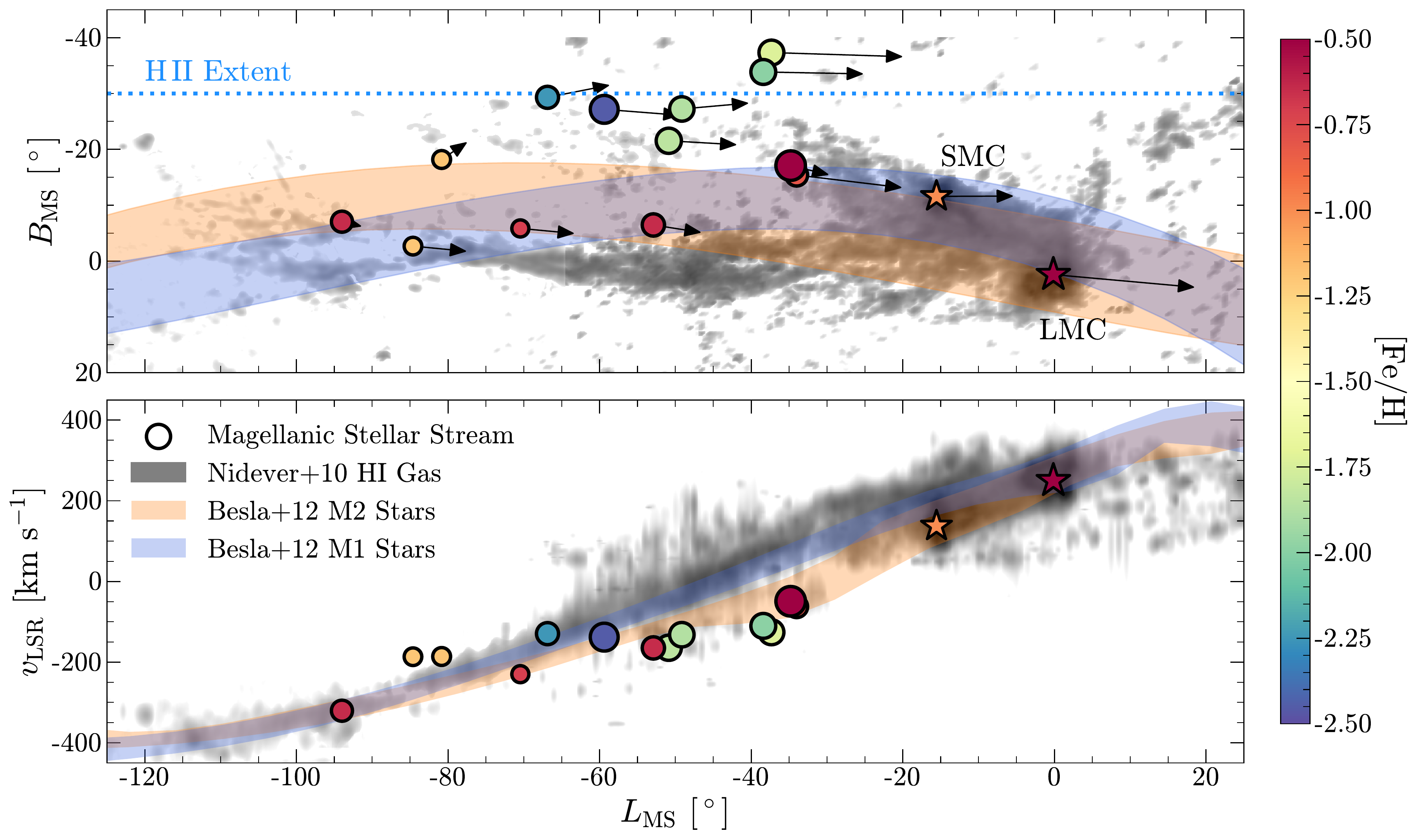}
    \caption{Comparing our MSS stars to the gaseous MS measurements from \cite{Nidever2010}, in the transverse stream coordinate \bms{} (top) and the LSR-corrected radial velocity (bottom). The log-scaled \ion{H}{1} column density of the MS is shown in gray, with our MSS candidates overlaid and colored by stellar metallicity. The point size is inversely proportional to the heliocentric distance. The dotted blue line indicates the approximate extent of the ionized \ion{H}{2} MS as measured by \cite{Fox2014}. Overlaid in blue (orange) is the predicted track of stripped MSS stars from Model 1 (Model 2) in \cite{Besla2012}. For the LMC and SMC colors we use median metallicities of $-0.5$ and $-1.0$ respectively, noting that they both have significant spreads in metallicity \citep{Hasselquist2021}. Arrows in the top panel denote the proper motion of the MSS stars and the Clouds, scaled to 50~Myr of motion.} 
    \label{fig:msgas}
\end{figure*}

To more directly compare our MSS stars to the gaseous MS, we transform our coordinates to the \cite{Nidever2008} MS frame, which defines a \lms{} coordinate aligned with the MS gas. 
In the top panel of Figure~\ref{fig:msgas}, the MSS stars are displayed over the total \ion{H}{1} column density of the MS \citep{Nidever2010}. 
The stars are colored by metallicity, and the point size is inversely proportional to the heliocentric distance --- the largest (smallest) point corresponds to 60 (120) kpc.
Proper motion vectors are overlaid, inflated to reflect $50$~Myr of motion. 
These vectors are corrected for the solar reflex motion. 
For the LMC and SMC we utilize the proper motion measurements of \cite{Kallivayalil2013} and \cite{Zivick2018}, respectively. 
We display on-sky tracks containing 90\% of simulation particles for both MSS models from \cite{Besla2012}.
The bottom panel of Figure~\ref{fig:msgas} shows the LSR-corrected radial velocity of the MS and our MSS candidates, again with the \cite{Besla2012} models overlaid.

Two distinct populations can be seen in our MSS sample shown in Figure~\ref{fig:msgas}: a spatially thin metal-rich component that closely traces the \ion{H}{1} gas and predicted MSS track from simulations, and an extended metal-poor component that is offset from the \ion{H}{1} gas by $\approx 20^\circ$. 
A distance gradient can be seen in the metal-rich component along the gas, with the MSS stars getting more distant further away from the Clouds (see Figure~\ref{fig:dgrad}). 
The metal-poor stars form a scattered cloud with little distance gradient, but still have velocities that closely trace the MS. 
Although only the \ion{H}{1} gas can be mapped out in detail and is consequently shown in Figure~\ref{fig:msgas}, the MS contains even more mass in the ionized \ion{H}{2} phase \citep[e.g.,][]{Lu1998, Fox2005, Fox2010, Fox2013, Fox2014, Fox2020}. 
Absorption line spectroscopy has demonstrated the presence of MS \ion{H}{2} gas out to \bms{}\,$\approx \pm 30^\circ$ (dashed blue line in Figure~\ref{fig:msgas}), resulting in a much more extended gas distribution. 
The \ion{H}{1} gas only traces the coldest, most dense phase of the MS, and the wider ionized MS encompasses most of the MSS stars presented here.

\begin{figure}
    \centering
    \includegraphics[width=\columnwidth]{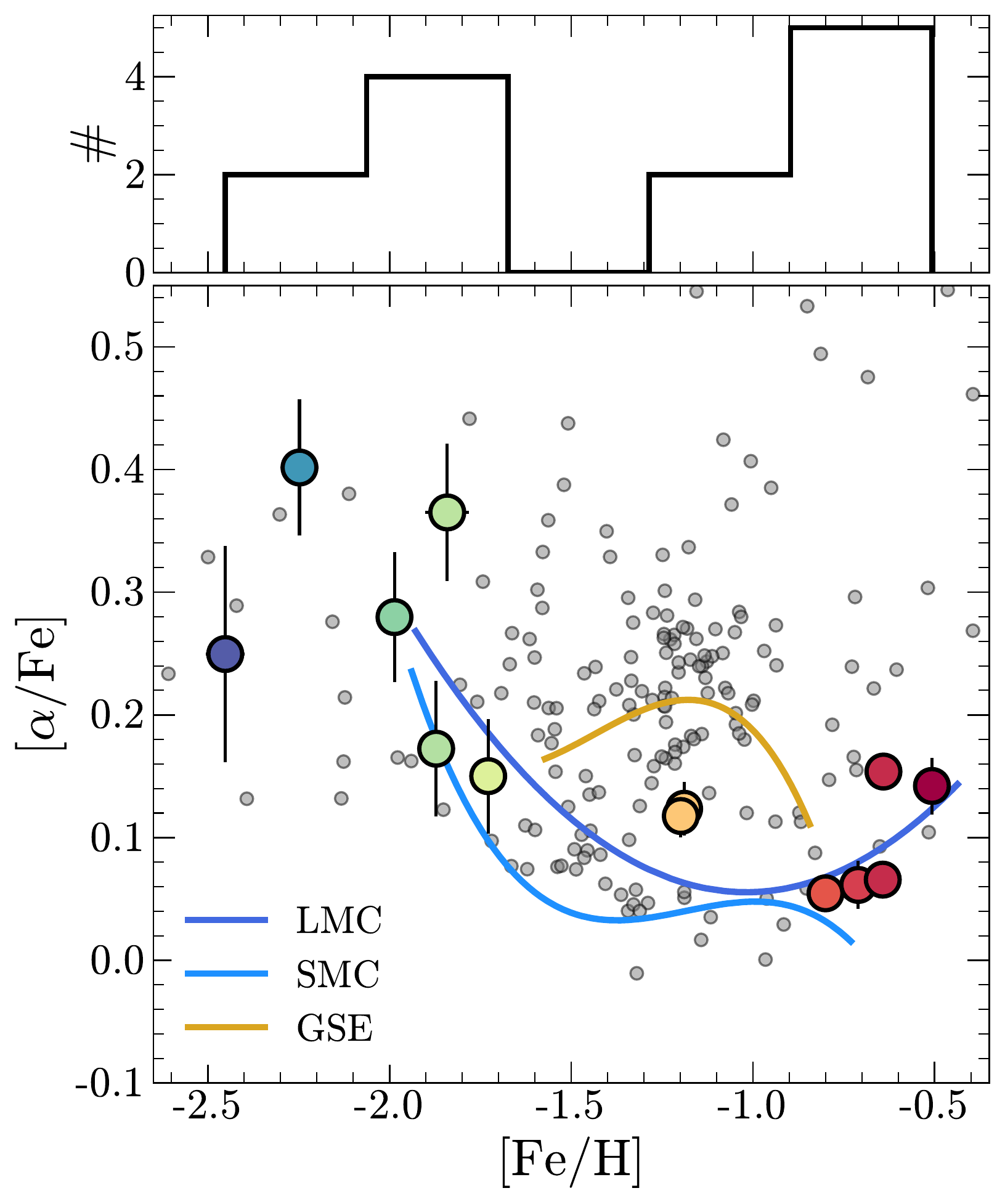}
    \caption{Kinematically-selected MSS stars in chemical abundance space, colored by metallicity as throughout this work. Median tracks for the LMC, SMC, and GSE dwarf galaxies are overlaid \citep{Hasselquist2021}.}
    \label{fig:tw}
\end{figure}

\subsection{Abundances and Distances}\label{sec:mss_dist}

Figure~\ref{fig:tw} shows the kinematically-selected stars in [Fe/H]--[$\alpha$/Fe] abundance space. 
We overlay the median chemical tracks for the LMC, SMC, and GSE from \cite{Hasselquist2021}. 
Since our MagE abundance scale uses the same spectroscopic models as the H3 Survey, we add 0.05~dex to the APOGEE [$\alpha$/Fe] tracks, informed by the offset measured in stars shared by APOGEE and H3. 
It is notable that all of the selected stars --- isolated purely based on kinematics and on-sky location --- lie along an alpha-poor abundance track that is consistent with formation in a recently-accreted dwarf galaxy with an extended star formation history, like the Clouds \citep{Font2006, Nidever2020}. 
This $\alpha$-deficient region of chemical space is also well-populated by non-MSS stars, as evidenced by the grey points in Figure~\ref{fig:tw}. 
This is predicted for an outer halo built out of relatively recently-accreted satellite galaxies \citep[e.g.,][]{Font2006}, and further work is underway to investigate the provenance of these stars. 

\begin{figure}
    \centering
    \includegraphics[width=\columnwidth]{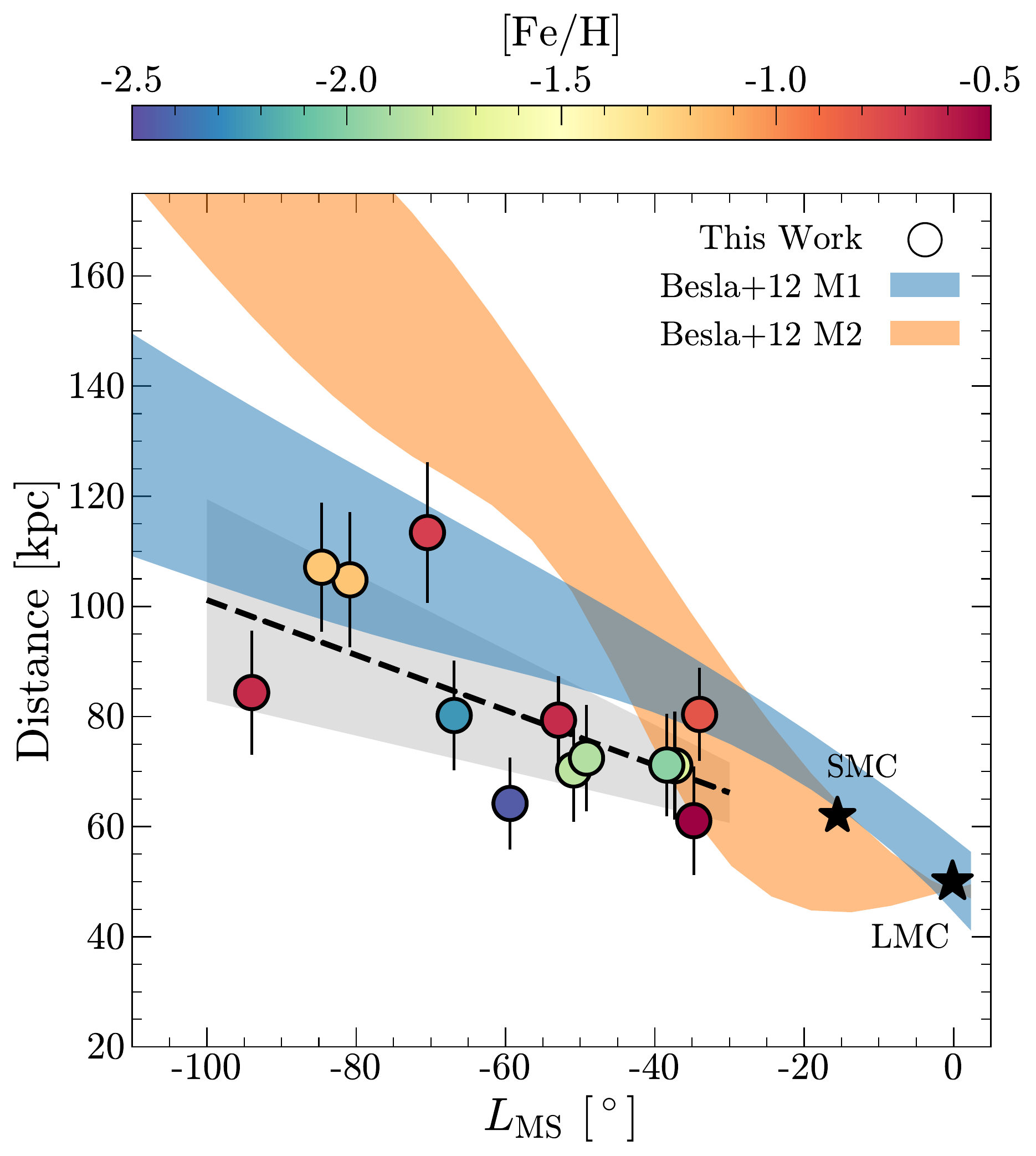}
    \caption{Spectroscopic distances for our identified MSS stars as a function of MS longitude. Shaded bands contain 90\% of the trailing tidal stellar debris from the simulations of \cite{Besla2012}. The dashed black line indicates the linear fit to the observed distance gradient along the MSS, along with a 1$\sigma$ shaded region.}
    \label{fig:dgrad}
\end{figure}

The observed distances of our MSS stars are presented versus \lms{} in Figure~\ref{fig:dgrad}, with points colored by metallicity on the same scale as Figure~\ref{fig:msgas}. 
We fit these stars with a linear distance trend --- taking into account heteroskedastic errors with a 10\% systematic error floor on individual distances --- and derive a distance gradient of \dgrad{}~kpc~deg$^{-1}$ along the \lms{} coordinate. 
If we instead only fit the 7 stars that directly trace the \ion{H}{1} MS with \bms{}$< 15^\circ$, the fitted distance gradient is almost identical. 
Later in $\S$\ref{sec:msgas} we use this observed distance gradient to re-compute the \ion{H}{1} mass of the MS. 

For comparison, we overlay onto Figure~\ref{fig:dgrad} trailing tidal debris from both MSS simulations in \cite{Besla2012}. 
Our identified MSS stars lie close to the predicted distances in the \cite{Besla2012} simulation, with a better match to Model 1 than Model 2. 
\cite{Lucchini2021a} utilized a different orbital orientation for the SMC compared to \cite{Besla2012}, and the trailing MS in their fiducial model is mostly contained within $20$~kpc. 
Our observed MSS stars all lie well beyond 60~kpc and are consequently inconsistent with this model. 
However, the more nearby MSS predicted by \cite{Lucchini2021a} cannot be entirely ruled out, since our spectroscopic sample was targeted towards the outer halo and does not contain more nearby stars. 
Furthermore, re-running the \cite{Lucchini2021a} model with different initial orbital configurations produces a wide range of MSS locations from $\approx 10-80$~kpc (S. Lucchini, private communication).

Regardless, our discovery of a substantial population of MSS stars from $60-120$~kpc --- coupled with their kinematic similarity to the simulations --- argues in favor of the LMC-SMC orbital configuration in \cite{Besla2012} as a plausible origin for these stars. 
Future spectroscopic surveys in the Southern hemisphere like SDSS-V \citep{Kollmeier2017} will fill in the nearer regions of the sky, and enable a search for observational analogs to the nearby MSS debris predicted by the fiducial model in \cite{Lucchini2021a}.

\begin{figure}
    \centering
    \includegraphics[width=\columnwidth]{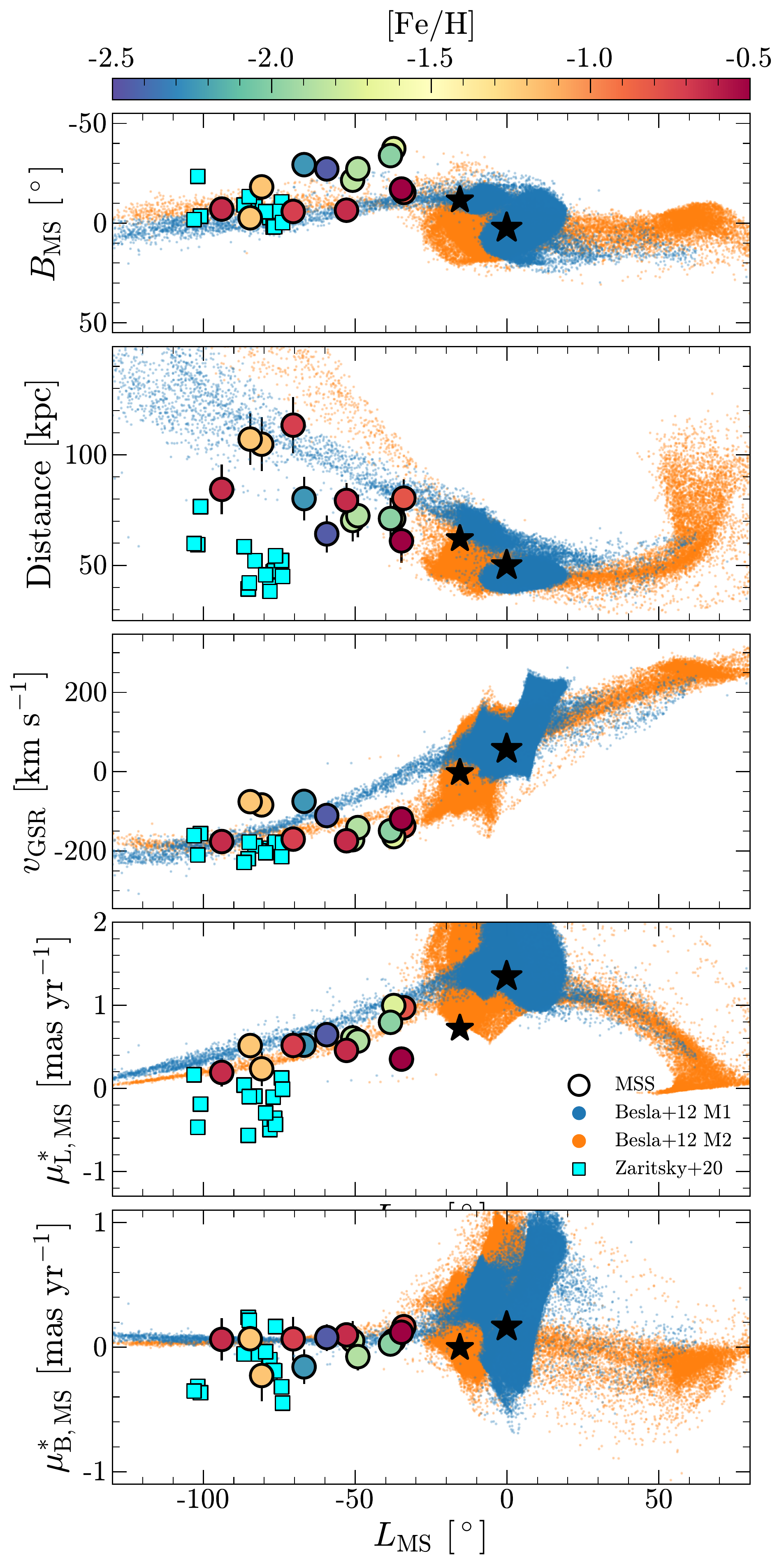}
    \caption{Projection of our MSS stars along the $L_\mathrm{MS}$ coordinate, against $B_\mathrm{MS}$ (panel 1), heliocentric distance (panel 2), Galactocentric radial velocity (panel 3), and proper motions along and transverse to the MS (panels 4 and 5). Particles from both tidal debris models in \cite{Besla2012} are shown, along with the Clouds-associated debris from the H3 Survey presented by \citetalias{Zaritsky2020}. 
    }
    \label{fig:comparison}
\end{figure}

\newpage

\subsection{Comparison to Prior Discoveries}

Figure~\ref{fig:comparison} summarizes various properties of our MSS stars along the \lms{} coordinate. 
We overlay the two Magellanic tidal debris models from \cite{Besla2012}.
To reiterate, Model 1 is designed to best match the velocity of the MS, whereas Model 2 presents a better match to the kinematics of the Clouds themselves. 
Our MSS stars are remarkably consistent with both models, exhibiting similar trends along \lms{} in \bms{}, heliocentric distance, Galactocentric radial velocity, and proper motions. 
We emphasize that these simulations were developed with no a priori knowledge of a Magellanic stellar stream, and were only designed to match the gas properties of the MS, and the kinematics of the Clouds. 
Regardless, it is clear that the \cite{Besla2012} simulations already reproduce the broad properties of our observed MSS. 
The key difference between our data and the \cite{Besla2012} models is the secondary diffuse population of metal-poor stars most clearly visible in Figure~\ref{fig:msgas}. 

We also show in Figure~\ref{fig:comparison} the Clouds-associated debris discovered in the H3 Survey by \citetalias{Zaritsky2020}. 
As discussed in that work, these debris provide an excellent match to the \cite{Besla2012} models in on-sky location and radial velocities. 
Furthermore, these stars lie at the mean metallicity of the SMC, and are systematically alpha-deficient compared to the field halo. 
However, there is a clear mismatch between the distances of the \citetalias{Zaritsky2020} stars and the \cite{Besla2012} models in the same region of sky, by over a factor of 2. \cite{Chandra2023} discussed the possibility that the \citetalias{Zaritsky2020} stars could be a part of the distant GSE debris identified in that work, but no definitive association could be made. 

The other dominant contributor to the halo at these distances is the Sagittarius Stream. 
Our identified MSS stars have markedly different properties than the Sagittarius Stream, particularly in distance and in proper motion along the \lms{} coordinate. 
The stars identified by \citetalias{Zaritsky2020} lie much closer on-sky to the MS than Sagittarius, although they have distances, Galactocentric radial velocities, and proper motions that are similar to Sagittarius. 
However, they do not fall in the $L_\mathrm{Z} - L_\mathrm{Y}$ locus that is characteristic of Sagittarius stars (see the bottom-left panel of Figure~\ref{fig:simL}; \citealt{Johnson2020a, Penarrubia2021}). 

Given their chemical abundances and coherent kinematics, the stars presented by \citetalias{Zaritsky2020} most plausibly also originate from the Clouds. 
One possibility is that they might have been stripped from a much earlier interaction between the Clouds, throwing these debris closer towards the MW. 
Another possibility is that these stars represent a tidally disrupted satellite that fell in with the Clouds, although their SMC-like metallicity argues against formation in a smaller satellite galaxy. 
It will be interesting to see if future N-body simulations of the MW-LMC-SMC system can reproduce both the \citetalias{Zaritsky2020} debris and the two-component MSS presented here, or if including an additional component like a satellite galaxy provides a better match. 

\subsection{Orbital Histories}\label{sec:orbits}

\begin{figure*}
    \centering
    \hspace*{0.5cm}\includegraphics[width=0.95\columnwidth]{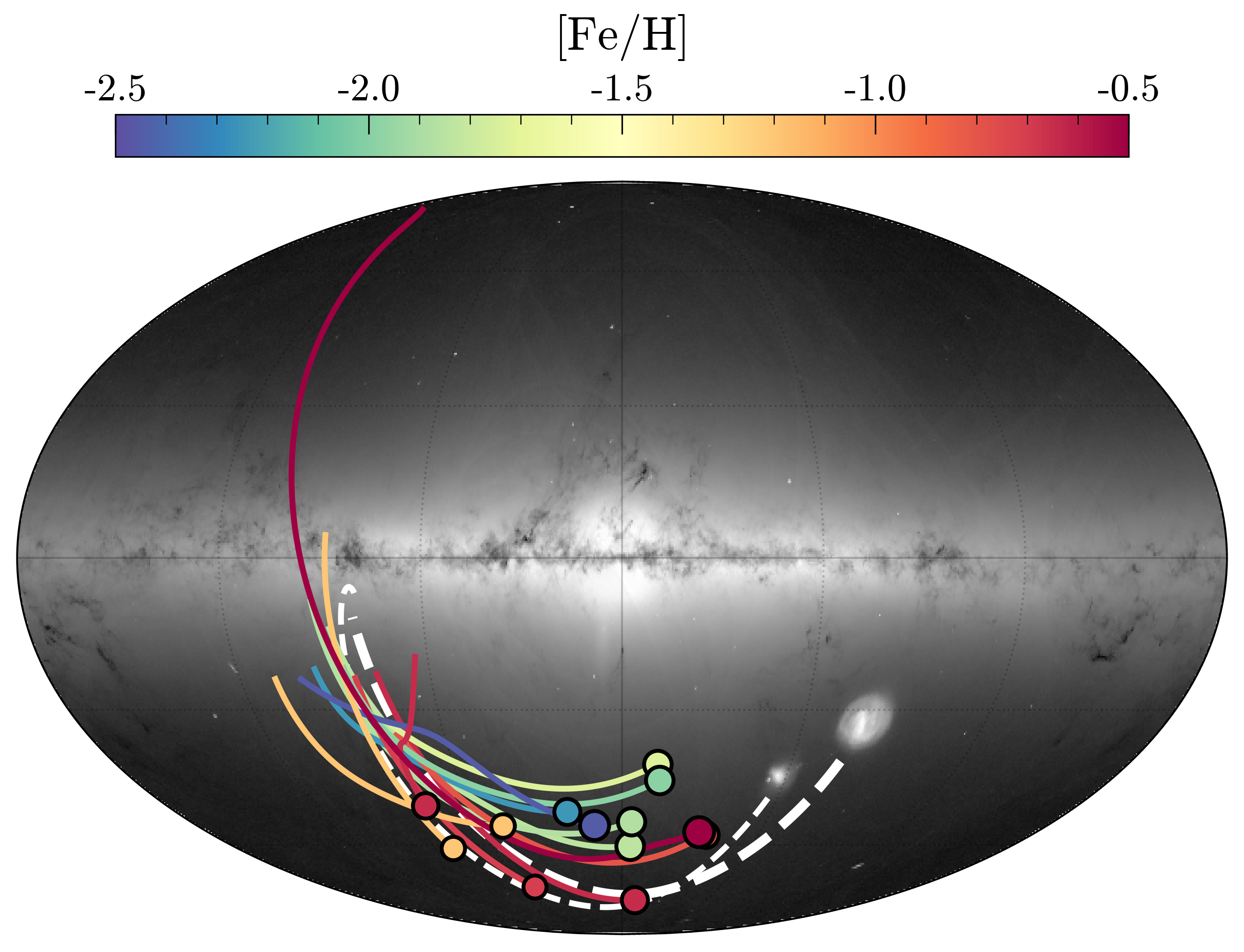}
    \includegraphics[width=\columnwidth]{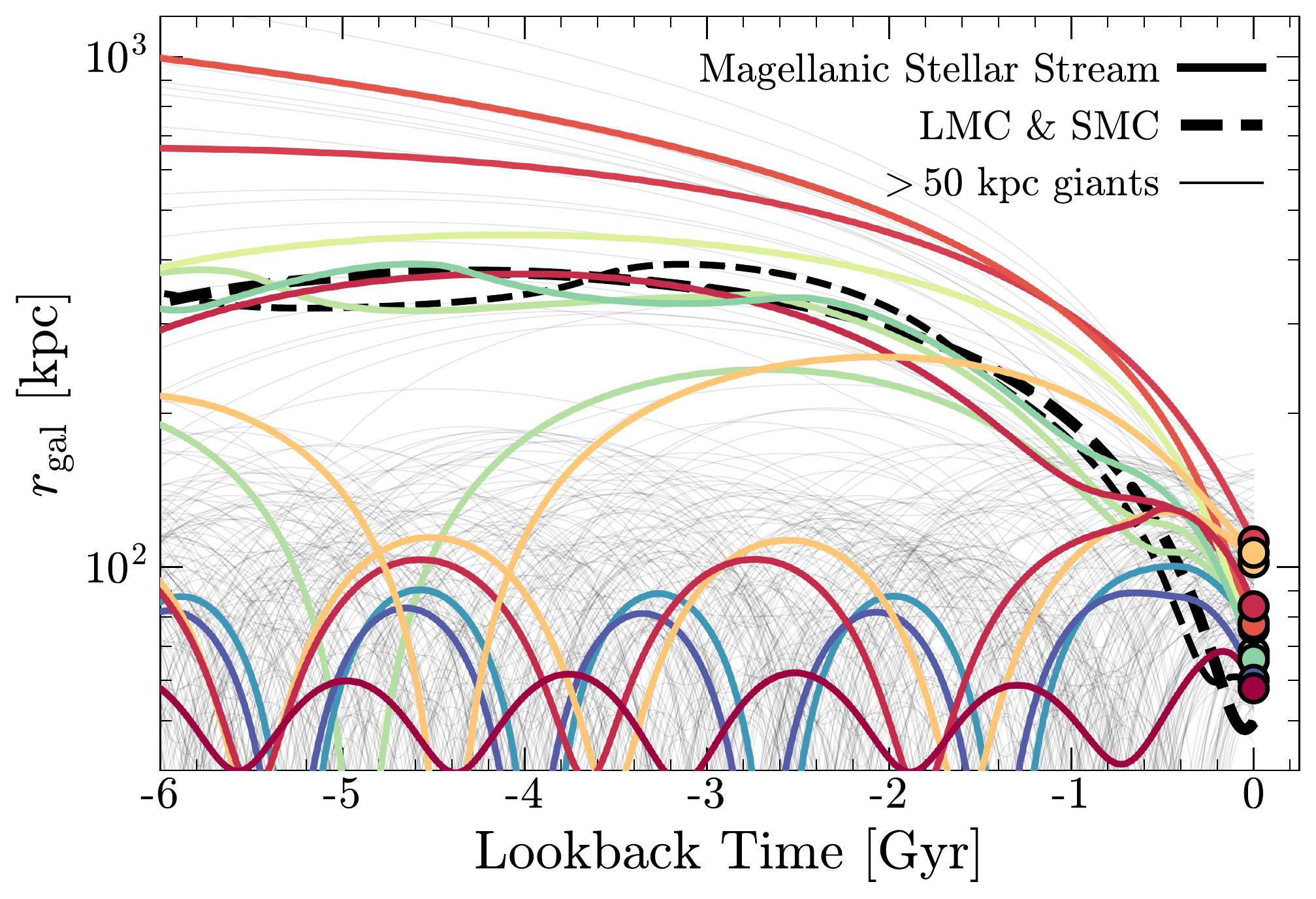}
    \includegraphics[width=\columnwidth]{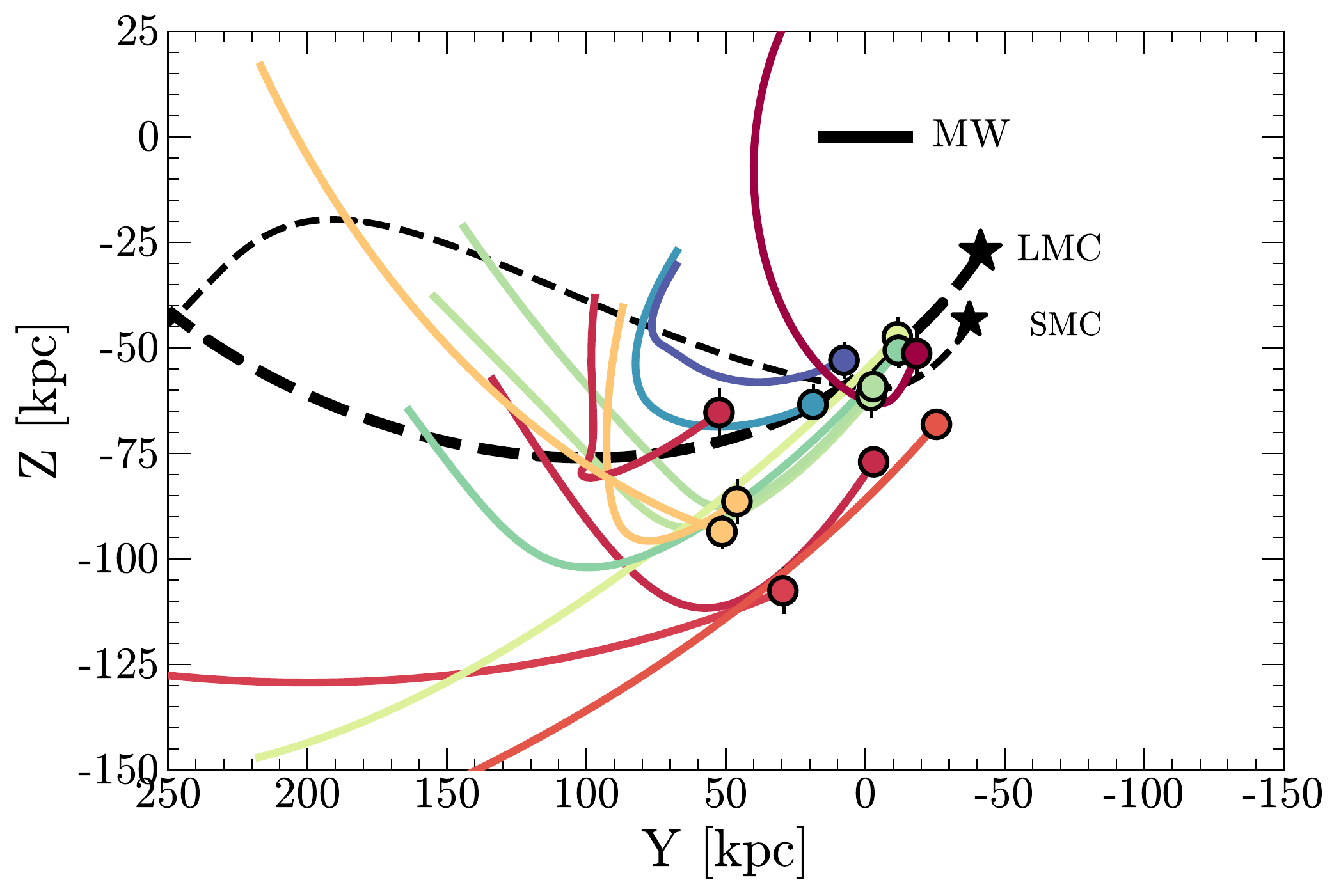}
    \includegraphics[width=\columnwidth]{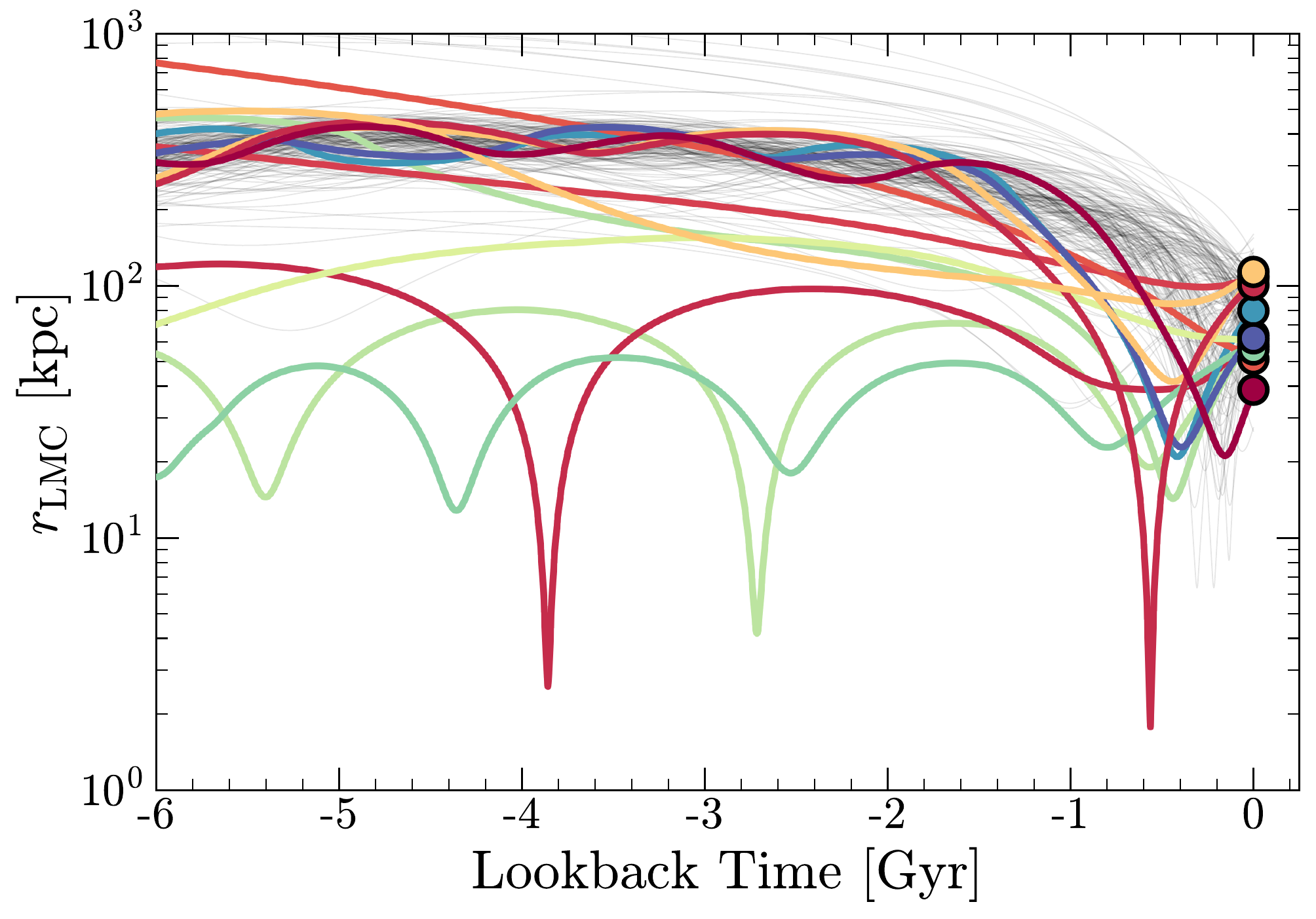}
    \caption{Integrated past orbits for our sample of MSS stars, in a rigid time-varying potential with LMC and SMC contributions adopted from \cite{Patel2020}. The top-left panel shows orbits for the past Gyr in Galactic coordinates and the bottom-left panel shows the same in the Galactocentric Y-Z plane. The top (bottom) right panel shows distances with respect to the MW (LMC) center-of-mass over 6~Gyr, for both our MSS stars and the entire $>50$~kpc giant sample from our survey. These panels only show orbits computed with mean phase-space positions, so several stars `unbound' to the LMC here have a non-zero probability of becoming bound in our orbital simulations (see Figure~\ref{fig:fbound}). 
    }
    \label{fig:orbits}
\end{figure*}

We have kinematically identified MSS stars using their outlying behavior in the $L_\mathrm{X}$ angular momentum (Figures~\ref{fig:simL},\ref{fig:mss}). 
Such selections based on angular momenta are straightforward and useful since they do not require orbital integration in an assumed potential, but yet strongly link outer halo debris to their progenitors (see, e.g., \citealt{Johnson2020a} and \citealt{Penarrubia2021} for an analogous selection of Sagittarius Stream stars). 
However, it is informative to explore the past orbits of these stars, to compare their past trajectories with those of the Clouds. 
Similar methods have been used to identify satellite galaxies that came in with the Clouds \citep{Patel2020, Erkal2020a}. 

We implement a rigid time-varying MW+LMC+SMC potential in \texttt{gala} \citep{gala, adrian_price_whelan_2020_4159870}. 
We fix the orbits of the LMC and SMC `on-rails' using trajectories calculated by \cite{Patel2020}, which are designed to match the latest kinematic measurements of the Clouds \citep{Kallivayalil2013, Gomez2015, Patel2017a, Zivick2018}. 
In that work, the three bodies are self-consistently modelled with a prescription for dynamical friction, and the MW barycenter is allowed to shift in response to the in-falling Clouds (i.e., it has a reflex motion). 
\cite{Patel2020} explore two Milky Way (halo + disk + bulge), three LMC (halo + disk), and two SMC (halo) potential models with varying mass and size parameters. 
We adopt their fiducial \texttt{MW1\_LMC2\_SMC1} model, using trajectories for the MW, LMC, and SMC going back 6~Gyr in steps of 1~Myr. 
This model has a $1.8 \times 10^{11}\,M_\odot$ LMC dark matter halo, which is the dominant time-varying component of the potential \citep[e.g.,][]{Erkal2019a, Shipp2021, Vasiliev2023}. 

At each step of the integration, we build a composite potential from the MW, LMC, and SMC components, centred on their trajectories in the \cite{Patel2020} model. 
We emphasize that our time-varying potential is entirely rigid, and does not self-consistently model the deformation of the potentials.
We integrate the phase space positions of all of our 191 distant giants back in time for 1~Myr in the present-day potential snapshot, then update the potential to the next timestep, and so on up to 6~Gyr in the past. 

Figure~\ref{fig:orbits} illustrates the integrated past orbits of our kinematically-selected MSS stars. 
As the bottom-right panel of Figure \ref{fig:orbits} demonstrates, some of our MSS stars actually become bound to the LMC-SMC system in the past, strongly associating them with the Clouds. 
This does not mean the other stars were not bound to the Clouds in the past: these trajectories are very sensitive to uncertainties in the phase space position of the stars and Clouds. 
Furthermore, the simplistic and rigid time-varying potential does not factor in the range of other physical processes that would have influenced these orbits, including ram pressure, tidal forces, and deformation.
Regardless, it is compelling that a large fraction of our MSS stars trace directly back to the Clouds in this model (top-left panel of Figure~\ref{fig:orbits}). 

We have experimented with modifying the time-varying potential to the other orbital models presented by \cite{Patel2020}. 
In particular, there are two more LMC models that have a factor of $\approx 2$ lighter and heavier LMC than the fiducial model, respectively. 
We find that adopting a different LMC mass --- and self-consistently updating both the orbital trajectory and potential in our model --- qualitatively changes the trajectories in Figure~\ref{fig:orbits}, but does not dramatically alter the results. 
Future work should self-consistently model the deformation of the potentials to investigate the effect of our assumed rigid potentials. 

\begin{figure}
    \centering
    \includegraphics[width=\columnwidth]{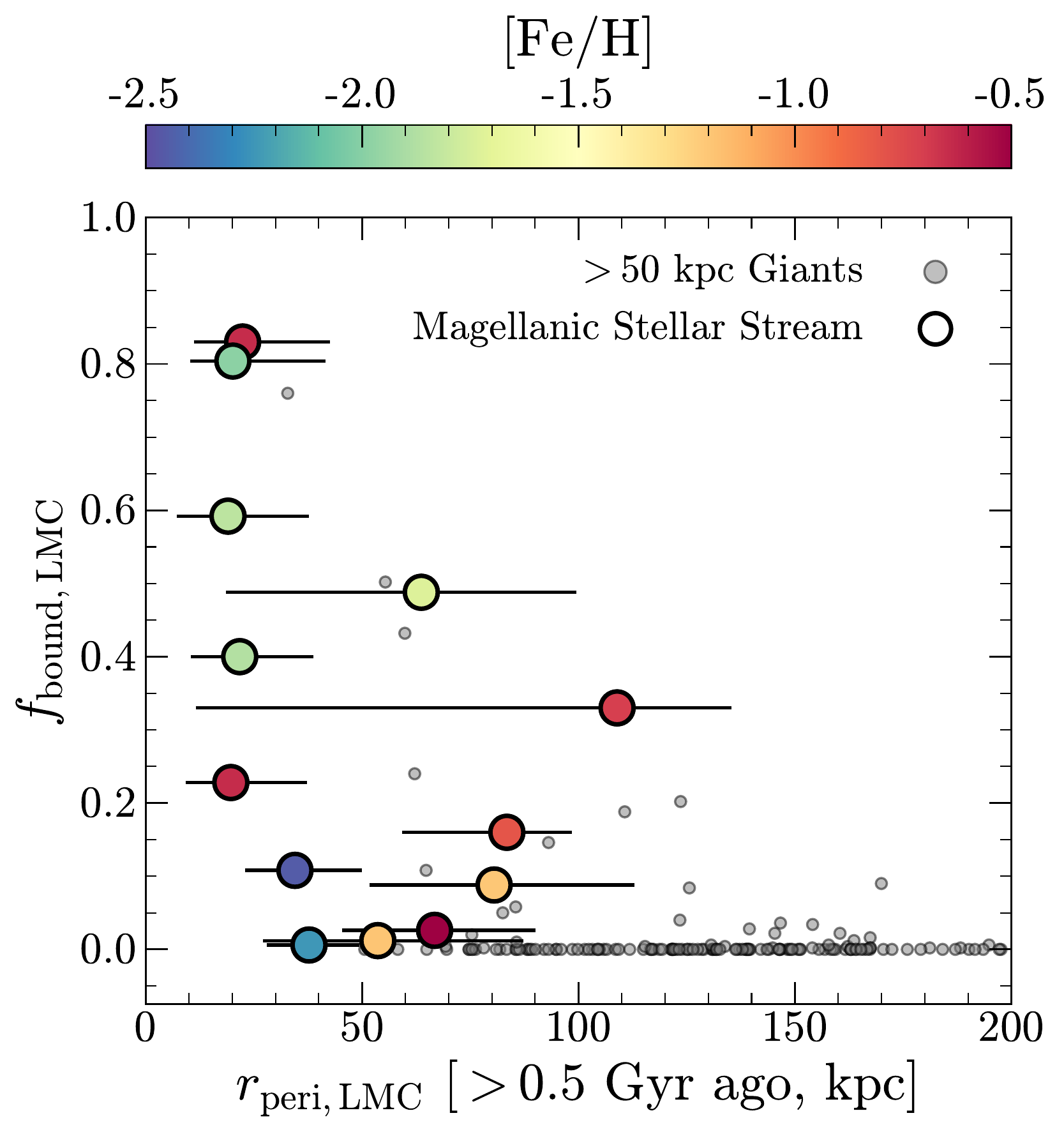}
    \caption{Results from our Monte Carlo trials to determine which stars in our $> 50$~kpc sample have orbits that plausibly lead back to the Clouds (see text for details). 
    Orbits are integrated backwards for 6~Gyr in the MW+LMC+SMC potential. 
    The horizontal axis shows the long-term (older than $0.5$~Gyr) orbital pericenter relative to the LMC, with error bars denoting the 16th and 84th quantiles. 
    The vertical axis shows the fraction of Monte Carlo trials in which the star remains bound to the LMC, quantified by the presence of a pericenter $> 2$~Gyr ago within the LMC tidal radius of 60~kpc. }
    \label{fig:fbound}
\end{figure}

To quantify whether our MSS stars have past orbital trajectories binding them to the Clouds, we perform a Monte Carlo experiment.
We sample the observed phase-space measurements within their Gaussian uncertainties, and integrate past orbits for 6~Gyr in the above described potential. 
In each trial, we check whether the star has a pericenter $2-6$~Gyr ago within $60$~kpc of the LMC --- if so, we consider it `bound' to the LMC in that trial. 
The threshold of $60$~kpc was chosen to roughly match the radius within which the LMC's potential dominates the MW's potential in the present day, and our results are insensitive to the exact threshold used. 
A parameter $f_\mathrm{bound, LMC}$ is consequently derived, which quantifies the fraction of a star's phase space distribution that leads to a past orbit bound to the LMC. 

Figure~\ref{fig:fbound} plots $f_\mathrm{bound, LMC}$ against the median pericenter relative to the LMC for our MSS stars, in addition to all stars in our $> 50$~kpc giant sample. 
Only pericenters older than 0.5~Gyr are considered, to remove trivial recent encounters as the LMC moves through the field of halo stars. 
Our simplistic angular momentum selection in Figure~\ref{fig:mss} isolates all stars with $f_\mathrm{bound, LMC} \gtrsim 0.3$, validating the use of angular momenta to find stars that move with/from the Clouds. 
Four of our MSS stars have $f_\mathrm{bound, LMC} \lesssim 0.1$. Since all other evidence points to them being plausible stream members, this likely represents a shortcoming in our simplistic orbital model --- both the adopted rigid potential and past Cloud orbits are uncertain in detail. 

Conversely, there are a couple of stars in our MagE sample with significant $f_\mathrm{bound, LMC}$ that were not selected as `MSS' members by our angular momentum cuts (gray points with $f_\mathrm{bound, LMC} \gtrsim 0.5$ in Figure~\ref{fig:fbound}). 
Examining these stars in detail, they are far from the MS on the sky, and have distances and/or kinematics that are quite discrepant from any simulation-based expectations of the MSS. 
A plausible explanation for these high $f_\mathrm{bound, LMC}$ stars is that they are field halo stars that have been stirred up by the LMC, or simply have orbits that coincidentally lead back to the LMC. 
In Appendix~\ref{sec:fieldhalo}, we estimate the magnitude of this latter effect using a mock smooth outer halo, and demonstrate that these chance alignments are rare ($\lesssim 4\%$ of field halo stars) would not artificially create the MSS we present in this work. 

\newpage

\subsection{Mass of the Magellanic Stellar Stream}

Our MSS sample consists of \nmss{} stars with $\log{g} < 1.5$. 
We can use this to place a lower limit on the stellar mass contained in the MSS, by integrating the expected unseen stellar mass given the number of stars we observe. 
This is a relatively weak lower limit for two reasons: first, our spectroscopic survey of the outer halo is far from complete in this $\log{g} < 1.5$ range at these distances. 
Our survey strategy aims to observe all stars in our sample beyond $100$~kpc, but only sparsely samples the $50-100$~kpc range. 
Furthermore, within this survey sample, we have aimed to produce a pure sample of MSS stars rather than a complete one. 

We utilize a 10~Gyr, [Fe/H]~$= -1.5$ MIST isochrone \citep{Choi2016}, and integrate the stellar mass with $\log{g} > 1.5$ using a \cite{Kroupa2001} IMF. 
We normalize this `visible' mass to be equal to the expected mass of our \nmss{} observed stars, and consequently integrate the total stellar mass of this simple stellar populations. 
This results in a lower bound on the stellar mass of the MSS $M \gtrsim 10^{4.5}\,M_\odot$. 
The tidal models of \cite{Besla2012} predict of order $M \gtrsim 10^{6}\,M_\odot$ of stars in the trailing arm of the MSS. 
Although this simulation-based prediction is highly uncertain, it remains likely that we have only unearthed a small portion of the observable giant stars in the MSS. 

\subsection{Mass of the H I Magellanic Stream}\label{sec:msgas}

\begin{figure}
    \centering
    \includegraphics[width=\columnwidth]{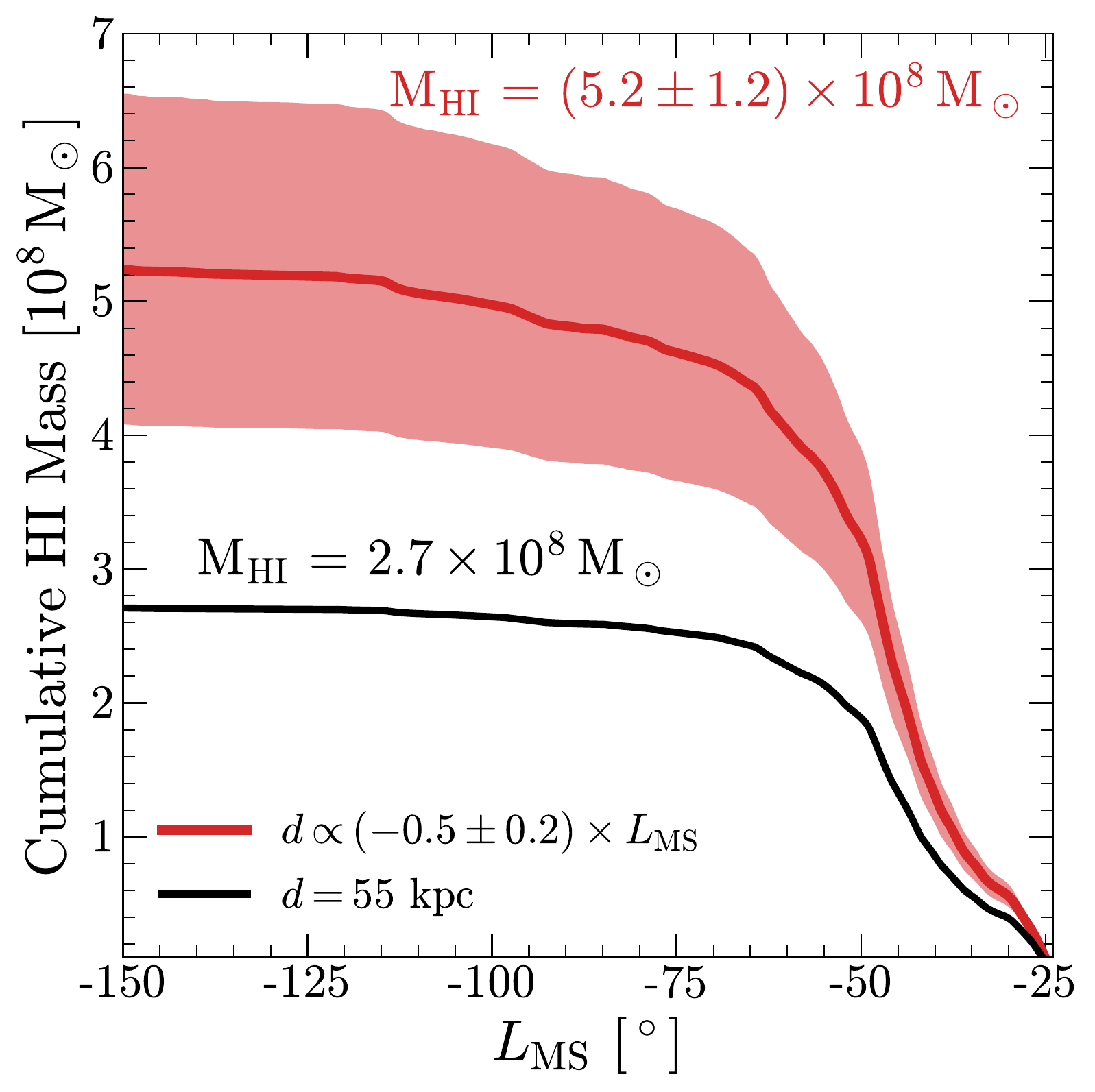}
    \caption{Integrated mass in the neutral MS gas, for a fixed distance of 55~kpc (black) and with our measured distance gradient for the stellar stream (red). The shaded band encompasses the 1$\sigma$ uncertainty on our distance gradient.}
    \label{fig:gasmass}
\end{figure}

Mass estimates of the gaseous MS require a distance to convert the \ion{H}{1} column density to a total mass. 
Fiducial estimates place the stream's \ion{H}{1} mass at $2.7 \times 10^8\,(d^2/55)\,M_\odot$, where $d$ is in kpc, and is chosen to match the mean distance of the Clouds \citep{Bruns2005,D'Onghia2016}. 
For comparison, the main body of the LMC contains an \ion{H}{1} mass $\approx 4.4 \times 10^8\,M_\odot$ \citep{Bruns2005}. 
In $\S$\ref{sec:mss_dist}, we measured a significant distance gradient with our MSS stars, \dgrad{}~kpc~deg$^{-1}$. 
It is reasonable to expect that the MS gas approximately follows the stellar component in distance gradient \citep[e.g.,][]{Nidever2010, Besla2012,Besla2013}. 
Here we re-derive the \ion{H}{1} mass of the MS, assuming that our stellar distance gradient applies to the gas as well. 

We utilize the HI MS maps from \cite{Nidever2010}, selecting a region with $< -150^\circ < L_\mathrm{MS} < -24^\circ$ such that the integrated mass matches the fiducial estimate of $2.7 \times 10^8\,(d^2/55)\,M_\odot$. 
This region includes the so-called `interface region' between the Clouds and MS, but excludes gas from the LMC and SMC themselves. 
We re-integrate the mass in the gaseous MS out to $L_\mathrm{MS} = -150$ with a fixed distance, and then with our measured distance gradient, and display the results in Figure~\ref{fig:gasmass}. 
The shaded red band indicates our new measurement, and the corresponding $1\sigma$ uncertainty propagated from the uncertainty in the slope of the distance gradient. 
We infer an \ion{H}{1} mass of \gasmass{} for the trailing MS, about a factor of two larger than fiducial estimates at fixed distance. 

This new value implies that the MS has an \ion{H}{1} gas mass comparable to or even greater than the LMC today ($\approx 4.4 \times 10^8\,M_\odot$, \citealt{Bruns2005}), albeit still less than the estimated ionized \ion{H}{2} gas in the MS ($\sim 10^9\,M_\odot$, \citealt{Fox2014}). 
The MS is the dominant contributor to the MW's gas inflow rate, and will perhaps trigger an increase in the Galactic star formation rate if it survives its journey towards the disk \citep{Richter2017, Fox2019}. 
Pinning down the distance and consequently mass of the MS --- by identifying more MSS stars with reliable distances, and quantifying any expected offsets between the MS and MSS distances --- therefore has broad implications for the future evolution of our Galaxy. 

\begin{deluxetable*}{ccccccccc}
\label{tab:spec_mem}
\tablecaption{13 high-confidence members of the Magellanic Stellar Stream from our MagE survey.}
\tablehead{\colhead{\textit{Gaia} Source ID} & \colhead{RA} & \colhead{Dec.} & \colhead{$g$} & \colhead{$v_{r,\mathrm{GSR}}$} & \colhead{[Fe/H]} & \colhead{[$\alpha$/Fe]} & \colhead{Distance} & \colhead{$f_\mathrm{bound, LMC}$}\\ \colhead{EDR3} & \colhead{deg} & \colhead{deg} & \colhead{mag} & \colhead{$\text{km\,s}^{-1}$} & \colhead{dex} & \colhead{dex} & \colhead{kpc} & \colhead{---}}
\startdata
6611869850296545024 & 333.05859 & -33.8833 & 17.1 & -110.3 & $-2.45\pm0.05$ & $0.25\pm0.09$ & $64\pm5$ & 0.11 \\
6619464348908375808 & 329.91349 & -27.49912 & 17.6 & -74.7 & $-2.25\pm0.04$ & $0.40\pm0.06$ & $80\pm6$ & 0.01 \\
6465479360347429888 & 321.90473 & -51.3359 & 17.2 & -148.7 & $-1.99\pm0.05$ & $0.28\pm0.05$ & $71\pm6$ & 0.80 \\
6568195388720305280 & 333.31103 & -42.98636 & 17.4 & -140.8 & $-1.87\pm0.05$ & $0.17\pm0.06$ & $72\pm6$ & 0.40 \\
6544712569129986944 & 340.89748 & -41.18518 & 17.5 & -171.2 & $-1.84\pm0.06$ & $0.37\pm0.06$ & $70\pm6$ & 0.59 \\
6477086664083654528 & 316.1877 & -51.15244 & 17.2 & -163.6 & $-1.73\pm0.04$ & $0.15\pm0.05$ & $71\pm7$ & 0.49 \\
2633051523542392320 & 353.06779 & -4.94589 & 17.6 & -75.5 & $-1.20\pm0.02$ & $0.12\pm0.02$ & $107\pm5$ & 0.09 \\
2601407505880573696 & 339.05471 & -12.89394 & 17.7 & -83.7 & $-1.19\pm0.03$ & $0.12\pm0.02$ & $105\pm6$ & 0.01 \\
6495806880338282496 & 353.0589 & -56.37423 & 16.8 & -136.9 & $-0.80\pm0.02$ & $0.05\pm0.01$ & $80\pm3$ & 0.16 \\
2392166457384504320 & 354.26508 & -19.37407 & 17.9 & -169.9 & $-0.71\pm0.03$ & $0.06\pm0.02$ & $113\pm6$ & 0.33 \\
2311053770911989504 & 359.37789 & -36.24547 & 16.7 & -174.1 & $-0.64\pm0.01$ & $0.15\pm0.01$ & $79\pm1$ & 0.83 \\
2661637550258741376 & 346.0394 & 2.60228 & 17.7 & -176.0 & $-0.64\pm0.03$ & $0.07\pm0.01$ & $84\pm7$ & 0.23 \\
6499025043497382272 & 349.88113 & -55.86038 & 17.9 & -119.1 & $-0.51\pm0.04$ & $0.14\pm0.02$ & $61\pm8$ & 0.03
\enddata
\tablecomments{Uncertainties are purely statistical, as returned by the \texttt{MINESweeper} fitting routine. $f_\mathrm{bound, LMC}$ denotes the fraction of Monte Carlo orbit trials in which the star becomes bound to the LMC $\gtrsim 2$~Gyr ago (see $\S$\ref{sec:orbits}).}
\end{deluxetable*}

\section{Discussion}\label{discuss}

\subsection{The Magellanic Stellar Stream}

We have presented the discovery of \nmss{} high-confidence members of the Magellanic Stellar Stream (MSS), the elusive stellar counterpart to the gaseous Magellanic Stream (MS). 
These stars have spectroscopic distances placing them between $50-120$~kpc from the Sun, with a significant distance gradient away from the Clouds. 
Assuming the stars broadly trace the past orbit of the Clouds --- and consequently the approximate location of the gas --- a key implication of our discovery is that the MS itself extends out to $\gtrsim 100$~kpc and beyond, as has long been speculated \citep[e.g.,][]{Connors2006, Besla2007,Besla2012, Besla2013, Nidever2010}. 

With the tip of the MS extending out to $\sim 150$~kpc, the MS could serve as a powerful tool to constrain the Milky Way's gravitational potential at large distances. 
The key remaining uncertainty is linking the observed stellar distances to the implied gas distances. 
While updated simulations that reproduce both the MS and MSS will provide some intuition about any expected offsets between the gas and stars, it would be desirable to measure the distance to the gas directly. 
One promising avenue is a more densely sampled spectroscopic survey of the outer halo, using the efficient techniques outlined here to select distant K giants (see also \citealt{Conroy2018, Chandra2023}). 
Searching for MS absorption along lines of sight to giants at different distances would enable us to tomographically reconstruct the MS gas distribution in 3D, mapping its full spatial extent. 
Improved distances to the stellar component of the MS would be valuable as well, for which future southern surveys of RRL and BHB stars will be crucial. 

Our MSS stars can be cleanly divided into two populations on the basis of their on-sky positions and metallicity (see Figures~\ref{fig:msgas},\ref{fig:tw}). 
About half of the stars reside in a metal-rich ([Fe/H]\,$\approx -0.8$) stream that closely traces the \ion{H}{1} MS gas. 
The other half form a diffuse cloud that is distinctly metal-poor ([Fe/H]\,$\approx -2.0$) and offset from the MS by $\sim 20^\circ$. 
Yet, both of these populations exhibit extreme kinematics and past orbits that associate them to the Clouds with high probability (Figure~\ref{fig:fbound}). 

The most likely origin of the metal-rich stream is that it forms the bona-fide stellar counterpart to the gaseous MS, and was formed with it.
Even disregarding their kinematic association, it would be difficult to explain the existence of 5 stars beyond 60~kpc at [Fe/H]\,$\gtrsim -0.8$ without invoking origin in a massive dwarf galaxy like the LMC. 
These stars are metal-rich enough that they may have formed in the LMC disk itself, and perhaps became unbound during a direct collision between the Clouds \citep[e.g.,][]{Mastropietro2005}. 
There is a remarkable kinematic correspondence between our observations and predictions from the simulations of \cite[][see Figure~\ref{fig:comparison}]{Besla2012}. 
While our MSS stars have velocities most consistent with Model 2 from \cite{Besla2012}, their distances are closer to Model 1. 
Reiterating that neither model reproduces the observed kinematics of the Clouds and the gaseous MS, there is a strong need for updated simulations.

The origin of the metal-poor cloud is more speculative, since a direct on-sky analog does not exist in existing simulations of the MSS (Figure~\ref{fig:msgas}). 
A plausible explanation is that the metal-poor cloud was thrown out of the SMC's outskirts during a past interaction/collision between the Clouds. 
The mean metallicity of the SMC was [Fe/H]~$\sim -1.0$ around 2.5 Gyr ago, when most tidal models suggest the MS began to form \citep[e.g.,][]{Pagel1998, Harris2004, Besla2010, Cignoni2013}. 
Therefore, our [Fe/H]\,$\approx -2.0$ MSS stars were either ejected from the main body of the SMC even earlier, or originate from the lower-metallicity outskirts of the SMC \citep{Carrera2008, Dobbie2014, Grady2021}. 
Given their scattered positions and distances, they might originate from the stellar halo of the SMC, but more modelling will be required to investigate this possibility. 

Another possible origin  --- for the metal-rich stream at least --- is that it formed out of the MS gas itself. 
The MS gas has two filaments with LMC-like and SMC-like metallicity measurements respectively \citep{Gibson2000, Richter2013, Fox2013}. 
Given evidence for recent star formation in the leading arm of the MS \citep{Price-Whelan2019, Nidever2019}, it is plausible that similar star formation events occurred in the trailing MS as well. 
If the MSS stars were indeed formed out of the MS gas itself, they are likely younger than our adopted age prior limit of $4$~Gyr for our stellar isochrone fits, implying they might be $10-20\%$ further than we estimate. 
Precise stellar age estimates of the trailing MSS will be vital to investigate this possibility. 
The subgiant branch of a $\sim 1$~Gyr stellar population at 100~kpc is around $G \approx 20.5$. 
This is just within the reach of future kinematic searches with the full-term \textit{Gaia} dataset, but likely too faint for wide-field spectroscopic surveys.

\subsection{A Dual Magellanic Origin for The Pisces Plume: A Stream and a Wake?}

\begin{figure*}
    \centering
    \includegraphics[width=1.5\columnwidth]{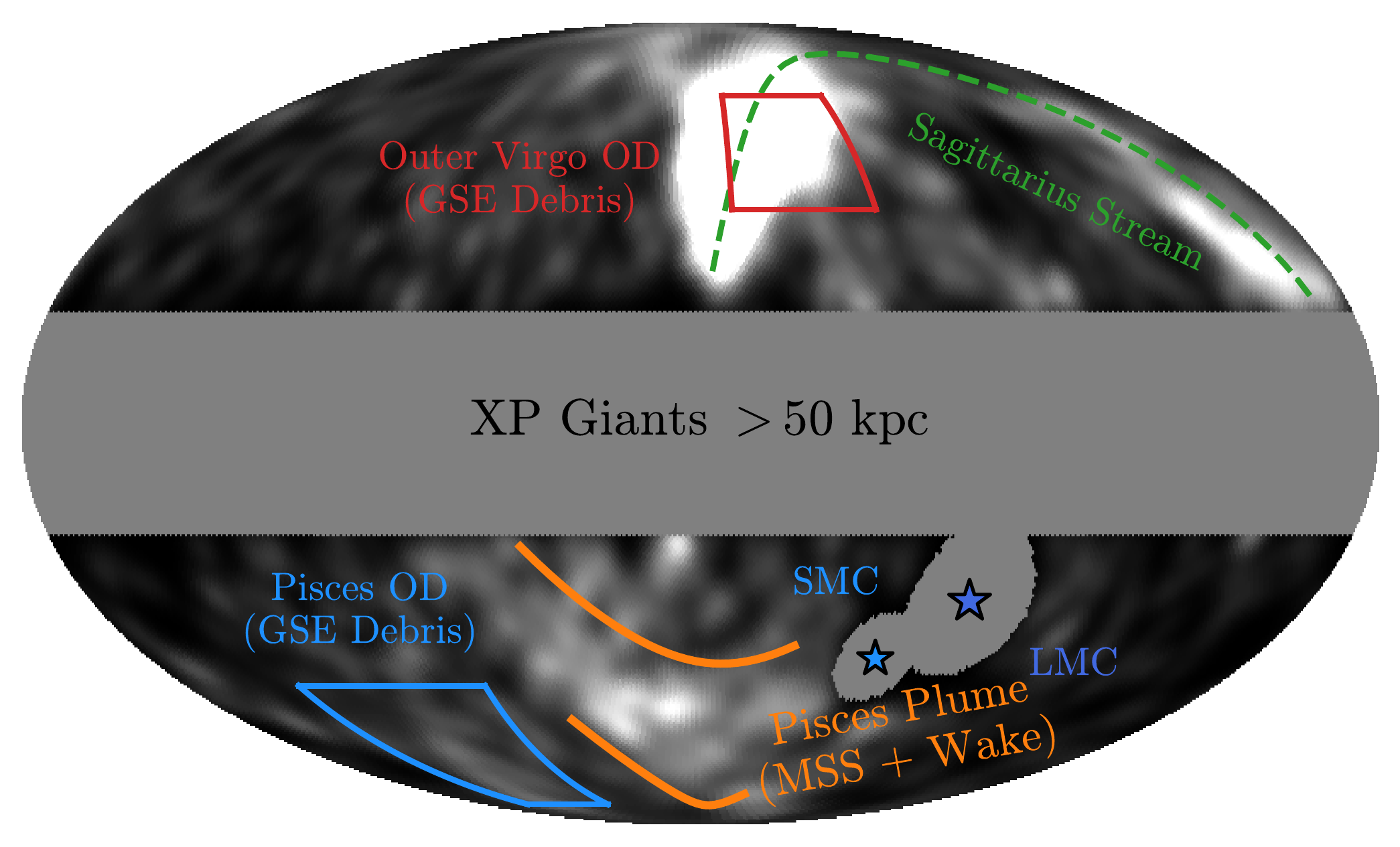}
    \caption{The emerging picture of the Milky Way's outer halo. 
    The colormap shows the density distribution of RGB stars beyond $> 50$~kpc from the \textit{Gaia} XP catalog of \cite{Chandra2023}, smoothed with a $5^\circ$ beam. 
    We highlight the prominent overdensities that dominate the outer halo, as well as their most plausible progenitors. 
    The Sagittarius Stream was first extensively mapped by \cite{Majewski2003}, and the Outer Virgo Overdensity \citep{Sesar2017} and Pisces Overdensity \citep{Sesar2007} were argued by \cite{Chandra2023} to be distant imprints of the GSE merger. 
    The elongated Pisces Plume was previously identified as the dynamical friction wake of the LMC \citep{Belokurov2019a, Conroy2021}.
    In this work, we have found a substantial population of MSS stars that overlaps with the Pisces Plume.}
    \label{fig:allsky}
\end{figure*}

Early maps of the $d \gtrsim 60$~kpc MW outer halo --- primarily with RRL stars --- had hinted at stellar overdensities in the distant southern sky towards the Pisces constellation \citep[the Pisces Overdensity or PO, e.g.,][]{Sesar2007,Watkins2009,Kollmeier2009,Sesar2010,Nie2015}. \cite{Chandra2023} used RGB stars with precise distances derived with \textit{Gaia} DR3 XP spectra to show that the Pisces Overdensity and Pisces Plume appear distinctly separated on-sky (Figure~\ref{fig:allsky}). 
Furthermore, \citet{Chandra2023} find that the kinematics and metallicities of stars in the PO suggest that at least some portion of the PO is composed of coherent debris from the Gaia-Sausage-Enceladus (GSE) merger, piling up near orbital apocenters. 

Armed with \textit{Gaia} DR2, \cite{Belokurov2019a} used all-sky RRL to uncover a plume-like elongation near the PO to larger distances, stretched along the same direction as the gaseous MS, suggesting an origin related to the Clouds. 
Based on the kinematics of this `Pisces Plume', they argued that it predominantly represented the dynamical friction wake imprinted on the MW's halo by the LMC's infall, rather than stripped stars from the Clouds. 
\cite{Conroy2021} utilized all-sky RGB stars to argue that the southern overdensity and a northern counterpart respectively correspond to the dynamical friction wake and `collective response' of the LMC's infall, matching predictions from simulations \citep{Garavito-Camargo2019,Garavito-Camargo2021}. 
However, they find a southern overdensity that is a factor of two stronger in the data than simulations predict, perhaps hinting at multiple populations in the Pisces Plume. 

Of the 191 stars beyond $50$~kpc analyzed in this work, 45 were targeted from the Gaia XP sample of \cite{Chandra2023} to lie within the Pisces Plume at $50\lesssim d \lesssim 100$~kpc. 
From the \nmss{} stars we identify as high-confidence members of the MSS, 7 were targeted from this Pisces Plume sample, with the remaining 6 originating from our broader $\gtrsim 100$~kpc selection. 
Therefore, at least $7/45$ or $\gtrsim 15\%$ of stars in the Pisces Plume appear to be confidently identified as debris from the Clouds. 
This is certainly a lower limit, since our MSS sample was designed towards purity rather than completeness, and several stars from the $\gtrsim 100$~kpc selection lie within the Pisces Plume on-sky but are too faint for the \textit{Gaia} XP sample. 

Up to $50\%$ of stars in the Plume selection lie along the [$\alpha$/Fe]-deficient track in chemical space (e.g., top right panel of Figure~\ref{fig:mss}), some of which may be MSS stars with less extreme $L_\mathrm{X}$ angular momenta. 
Furthermore, it is reasonable to expect a high fraction of the field halo at these distances to consist of stars on the [$\alpha$/Fe]-deficient chemical track, assuming the outer halo contains numerous disrupted dwarf galaxies. 
It will be valuable to disentangle the contribution of [$\alpha$/Fe]-deficient stars from larger progenitors like Sagittarius and the Clouds, to obtain a census of the smaller, more ancient dwarf galaxies that have merged and perished in the Galactic outskirts.  

In summary, although the MSS clearly contributes to the Pisces Plume, it is probably not the sole cause of this stellar overdensity. 
It remains likely that field halo stars are perturbed in the wake of the LMC due to dynamical friction, further enhancing the Pisces Plume in a region of sky coincident with the MSS. 
As aforementioned, \cite{Conroy2021} find an overdensity in the Plume that is a factor of two larger than that expected from models of the dynamical friction wake alone.
If up to $\approx 50\%$ of the Plume is indeed composed of stars from the MSS, this tension could be alleviated. 

Although we have here used tailored cuts to excise high-confidence members of the MSS, accurately measuring the relative contribution of both Clouds-induced progenitors of the Plume --- the stream and the wake --- will likely require a sophisticated and probabilistic chemodynamical model. 
This will be the subject of future work from our survey, once we have spectroscopically observed all our targets in this region of the sky. 

\section{Conclusions}

In this work, we have discovered a stellar counterpart to the gaseous MS as a part of our ongoing spectroscopic survey of the Galactic outskirts. Our key conclusions are summarized as follows: 

\begin{itemize}
    \item From a spectroscopic sample of 191 stars observed with MagE beyond $50$~kpc, we isolate \nmss{} high-confidence members of the MSS based on their extreme angular momenta that strongly associate them with the Clouds (Figures~\ref{fig:simL},\ref{fig:mss}).

    \item The LMC/SMC origin of these kinematically-selected stars is further supported by two direct observations: the stars closely trace the gaseous MS on-sky, and have chemical abundances consistent with their formation in the Clouds (Figures~\ref{fig:mss},\ref{fig:msgas},\ref{fig:tw}). 

    \item The MSS stars have past orbits that coincide with those of the Clouds. When integrated backwards in a rigid time-varying potential that includes the LMC and SMC, the majority of our MSS stars could have a non-zero probability of being bound to the LMC-SMC system in the past (Figures~\ref{fig:orbits},\ref{fig:fbound}). 

    \item About half of our MSS stars are relatively metal rich ($-1.2 \lesssim \feh{} \lesssim -0.5$) and form a thin stream right along the \ion{H}{1} MS gas, stretching up to a hundred degrees away from the Clouds. 
    The other half are metal-poor ($-2.5 \lesssim \feh{} \lesssim -1.7$), and form a diffuse cloud that is offset from \ion{H}{1} MS gas, but is still within the ionized \ion{H}{2} extent of the MS (Figure~\ref{fig:msgas}). 
    We argue that the thin stream is the bona-fide stellar counterpart formed with the gaseous MS, whereas the diffuse cloud was perhaps expelled from the SMC during an earlier interaction between the Clouds. 

    \item We measure a significant distance gradient in the MSS of \dgrad{}~kpc~deg$^{-1}$ along the \lms{} coordinate, with a mean distance of $\approx 80$~kpc (Figure~\ref{fig:dgrad}). 
    These measurements imply that the gaseous MS itself extends to $\gtrsim 100$~kpc and beyond, and its distance constrains the past orientation of the LMC-SMC system.
    
    \item We re-integrate the mass of the \ion{H}{1} MS gas under the assumption that the gas follows the distance gradient of the stars, obtaining \gasmass{} (Figure~\ref{fig:gasmass}). 
    This is a factor of two greater than fiducial mass estimates that assume the entire stream is at $55$~kpc, although more detailed simulations --- and perhaps direct distance measurements to the gas --- will be required to ascertain how closely the gas truly follows the stars. 

    \item Our MSS stars are spatially and kinematically consistent with past numerical simulations of tidally-stripped MSS stars (Figures~\ref{fig:msgas},\ref{fig:dgrad},\ref{fig:comparison}). 
    Interesting differences persist, chiefly the existence of the diffuse metal-poor population. 
    Our MSS stars will provide strong constraints on future simulations of the MW-LMC-SMC interactions, particularly via their 3D kinematics and distances. 
    Furthermore, their metallicities constrain where in the Clouds they might have been stripped from, with the metal-poor population almost certainly requiring formation in the outskirts of the SMC. 

    \item Our results suggest that at least $20\%$ and up to $50\%$ of the Pisces Plume outer halo overdensity is composed of debris from the Clouds.
    Accounting for this contribution would bring the observed Plume overdensity in line with simulations of the LMC's dynamical friction wake, but further work is required to perform a more quantitative separation between the MSS and field halo wake in this region. 
    
\end{itemize}

These are the first scientific results from our ongoing spectroscopic survey of the Milky Way's outer halo to 100~kpc and beyond. 
As Figure~\ref{fig:allsky} illustrates, full chemodynamical information is the key to revealing various progenitors of the richly structured outer halo. 
The Clouds play an outsize role in shaping these outskirts of our Galaxy, both by depositing stellar and gaseous debris, and via their strong gravitational influence on the stellar and dark matter halo. 
They were also the first Galactic satellites discovered, with records dating back to antiquity.
Armed with the novel constraints from the MSS discovered here, we can more directly trace the history of our Galaxy's most familiar and dominant neighbors. 

\appendix 

\section{MSS MagE Spectra}

\begin{figure}
    \centering
    \includegraphics[width=\columnwidth]{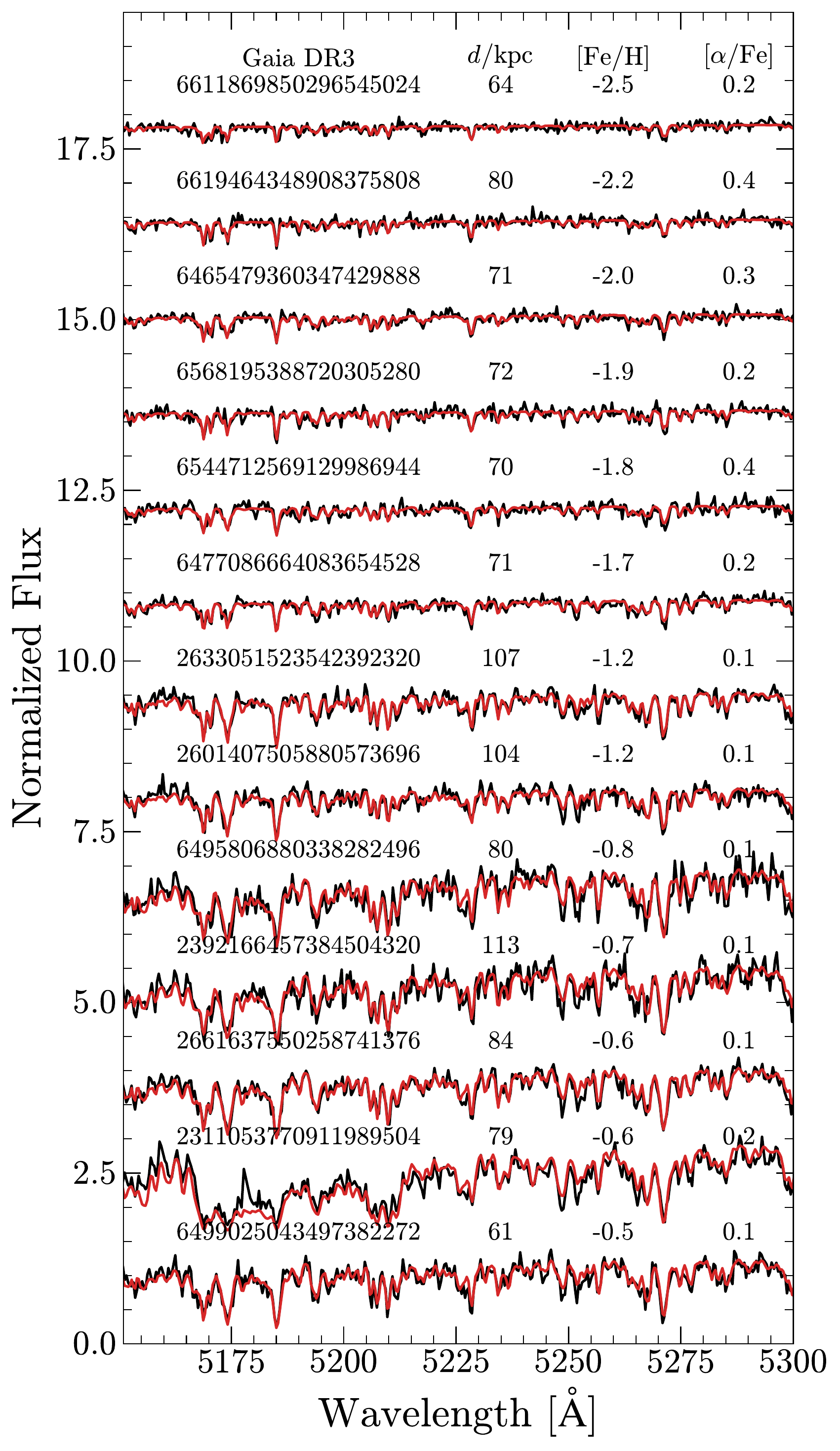}
    \caption{A segment of our MagE spectra for the MSS stars presented in this work, with the data shown in black and the fitted \texttt{MINESweeper} model shown in red. Spectra have been shifted to the rest frame.}
    \label{fig:mssspectra}
\end{figure}

In Figure~\ref{fig:mssspectra} we show a region of our MagE spectra for the \nmss{} high-confidence members of the MSS we identify in this work. For our \texttt{MINESweeper} fits to determine stellar parameters and abundances, we used a broader wavelength region from $4800-5500$~\AA{}.

\section{Ruling out False Positives From the Field Halo}\label{sec:fieldhalo}

In this work, we have kinematically selected a population of stars beyond 50~kpc with angular momenta and chemistry that are consistent with formation in the Clouds. Furthermore, integrating their orbits back in time results in a significant fraction becoming bound to the LMC (Figure~\ref{fig:fbound}). However, it is plausible to wonder whether such a population of stars might be a `false positive', artificially created by a tailored selection of field halo stars. The metal-rich, alpha-poor chemistry of the main stream stars beyond 60~kpc (up to [Fe/H]~$= -0.6$) already argues against this interpretation, pointing to formation in a metal-rich dwarf like the LMC. The remarkable correspondence between our selected stars and the tidal debris simulations of \cite{Besla2012} are also highly encouraging (Figure~\ref{fig:comparison}). Regardless, in this Appendix we repeat our selection on a mock stellar halo with a `smooth' distribution of star particles, to investigate the false positive rate of our selection. 

We select stars from the Gaia Universal Model Simulation (GUMS; \citealt{Robin2012}), a simulated catalog of Gaia measurements drawn from a smooth Milky Way based on the Besançon galaxy model \citep{Robin2003}. We select simulated giants with $50 < d/\mathrm{kpc} < 160$, $G < 19$, and which satisfy the angular momentum cuts shown in Figure~\ref{fig:mss}, resulting in 165 mock stars that match our kinematic selection. We then integrated their past orbits in the MW+LMC+SMC potential described in $\S$~\ref{sec:orbits}. Applying the same criteria to quantify whether a star becomes `bound' to the LMC --- a pericenter $> 2$~Gyr ago with $r_\mathrm{LMC} < 60$~kpc --- only 6/165 mock stars ($\approx 4\%$) do become bound. Half of these stars lie in the northern hemisphere, and only two have $L_\mathrm{MS} < 0$. They have on-sky positions, distances, and kinematics that are inconsistent with both the gaseous MS and the tidal models of \cite{Besla2012}. 

We therefore conclude that although our time-varying potential experiment can indeed result in a small fraction of field halo stars becoming `bound' to the LMC in the past, these stars would not look anything like the MSS we have identified. On the other hand, applying this kinematic selection to our data produces a stream with location and chemistry consistent with origin in the Clouds, cementing our confidence in their provenance. 

\begin{acknowledgments}

VC gratefully acknowledges a Peirce Fellowship from Harvard University. 
We thank Gurtina Besla, Himansh Rathore, and Hayden Foote for sharing their simulations of tidal MSS debris from \cite{Besla2012,Besla2013}.
We thank Ekta Patel for advice on our orbital integration scheme, and for sharing Cloud orbital trajectories from \cite{Patel2020}. 
VC is grateful to 
David Nidever,
Scott Lucchini,
Elena D'Onghia,
Andrew Fox,
Vadim Semenov, 
Turner Woody, 
and Gus Beane
for insightful conversations and feedback. 
We are indebted to the CfA and MIT telescope time allocation committees for enabling our long-term survey of the Galactic outskirts. 
We thank the staff at Las Campanas Observatory --- including Yuri Beletsky, Carla Fuentes, Jorge Araya, Hugo Rivera, Alberto Past\'en, Mauricio Mart\'inez, Roger Leiton, Mat\'ias D\'iaz, Carlos Contreras, and Gabriel Prieto --- for their invaluable assistance. 
RPN acknowledges support for this work provided by NASA through the NASA Hubble Fellowship grant HST-HF2-51515.001-A awarded by the Space Telescope Science Institute, which is operated by the Association of Universities for Research in Astronomy, Incorporated, under NASA contract NAS5-26555.
The H3 Survey is funded in part by NSF grant NSF AST-2107253.

The computations in this paper were run on the FASRC Cannon cluster supported by the FAS Division of Science Research Computing Group at Harvard University.
This work has made use of data from the European Space Agency (ESA) mission {\it Gaia} (\url{https://www.cosmos.esa.int/gaia}), processed by the {\it Gaia} Data Processing and Analysis Consortium (DPAC, \url{https://www.cosmos.esa.int/web/gaia/dpac/consortium}). Funding for the DPAC has been provided by national institutions, in particular the institutions participating in the {\it Gaia} Multilateral Agreement. 
This research has made extensive use of NASA's Astrophysics Data System Bibliographic Services.

\end{acknowledgments}

\software{\texttt{numpy} \citep{Harris2020}, 
\texttt{scipy} \citep{Virtanen2020}, 
\texttt{matplotlib} \citep{Hunter2007}, 
\texttt{gala} \citep{gala,adrian_price_whelan_2020_4159870}
\texttt{MINESweeper} \citep{Cargile2020}
}

\facilities{Magellan:Baade (MagE), Gaia, Sloan, PS1, 2MASS, WISE}

\bibliography{library,bib}
\bibliographystyle{aasjournal}

\end{document}